\documentclass[preprint]{aastex}

\received{}
\revised{}
\accepted{}
\ccc{}
\cpright{}{}

\newcommand{\kms}{km~s$^{-1}$}
\newcommand{\subsun}{\mbox{$_{\odot}$}}
\newcommand{\etal}{{\it et al.\/}}
\newcommand{\teff}{$T_{eff}$}
\newcommand{\grav}{log($g$)}

\newcommand{\ciso}{C$^{12}$/C$^{13}$}

\begin{document}

\title{C and N Abundances in Stars At the Base of the Red Giant 
Branch in M15
\altaffilmark{1} }

\author{Judith G. Cohen\altaffilmark{2},
Michael M. Briley\altaffilmark{3} and Peter B. Stetson\altaffilmark{4,5,6}}

\altaffiltext{1}{Based on observations obtained at the
W.M. Keck Observatory, which is operated jointly by the California
Institute of Technology, the University of California and the National
Aeronautics and Space Administration}

\altaffiltext{2}{Palomar Observatory, Mail Stop 105-24,
California Institute of Technology, Pasadena, California 91125
(jlc@astro.caltech.edu)}

\altaffiltext{3}{Department of Physics and Astronomy, University of Wisconsin
Oshkosh, 800 Algoma Boulevard, Oshkosh, WI 54901. Currently
on assignment to the National Science Foundation, 4201 Wilson
Boulevard,  Arlington, VA 22230. Any opinions, findings, conclusions 
and
recommendations 
expressed in this material are those of the author and
do not necessarily 
reflect the views of the National Science
Foundation.
(mike@maxwell.phys.uwosh.edu)}

\altaffiltext{4}{Dominion Astrophysical Observatory, 5071  West 
Saanich Road,
Victoria, British Columbia   V9E 2E7 Canada (Peter.Stetson@hia.nrc.ca)}

\altaffiltext{5}{Guest User,  Canadian Astronomy Data Centre, 
which is operated by the
Herzberg Institute of Astrophysics, National Research Council of Canada}

\altaffiltext{6}{Guest Investigator at the UK Astronomy Data Centre}

\begin{abstract}
We present an analysis of a large sample of
moderate resolution Keck LRIS spectra of subgiants and stars 
at the base of the RGB in the Galactic globular cluster 
M15 (NGC~7078),
most within the range $16.5 < V < 19.5 (1.2 < M_V < 4.2)$, with the
goal of deriving C abundances (from the G band of CH) and
N abundances (from the NH band at 3360\,\AA). Star-to-star
stochastic variations with significant range
in both [C/Fe] and [N/Fe] are found at all luminosities
extending to the subgiants at $M_V \sim+3$.

The C and N abundances appear anti-correlated, as
would be expected from the CN-cycle processing of stellar material.
Yet these M15 stars are considerably fainter than
the RGB bump, the point at which deep
mixing is believed to set in.  On this basis,
while the observed abundance pattern is consistent with 
proton capture nucleosynthesis,  we infer that
the site of the reactions is likely not
within the present sample.
The range of variation
of the N abundances is very large and the sum of C+N increases
as C decreases.  To reproduce this requires  
the incorporation not only of CN but also of ON-processed material.

Combining our work with that of \cite{trefzger83} for the brighter
giants in M15, we find strong evidence for additional
depletion of C among the most luminous giants.   This presumably
represents the first dredge up (with enhanced deep mixing) expected for 
such luminous cluster RGB stars in the course of normal
stellar evolution  as they cross the RGB bump.  

We compare the behavior of these patterns for C and N in globular clusters covering
a wide range of metallicity and of current mass.  While all clusters
studied show strong anti-correlated variations of C and N at all
luminosities probed,  the metal rich
clusters (M71, 47 Tuc and M5)
do not show evidence for the first dredge up among their most
luminous giants, while the metal poor ones (M5, M13, M92 and M15)
do.  Conversely, the metal poor clusters do not show evidence for the bimodality
in CH and CN line strengths seen in the metal rich clusters.

The collected data on C and N abundances in low luminosity GC stars
cannot be explained by the commonly invoked models for the chemical evolution of GC
stars; in particular ``pollution'' of existing low mass stars by ejecta
from intermediate mass AGB stars can be ruled out.  Pollution of
cluster gas by such stars prior to the formation of the lower mass stars we
observe today
can also be ruled out unless current models of nucleosynthesis and dredge up
into the surface layers of AGB stars are flawed; such models agree
qualitatively but disagree quantitatively with our data.
We are forced to assume that there was an extended period of star 
formation in GCs, and that a previous generation of more massive stars
evolved, ejected mass, and ``polluted'' with light elements the GC gas;
the low mass stars we see today formed afterwards.  A tentative scenario is developed
involving a initial phase of star formation heavily biased towards high
mass stars, with subsequent formation of intermediate, then low mass stars.

\end{abstract}


\section{Introduction\label{intro}}

By virtue of their large populations of coeval stars, the Galactic
globular clusters present us with a unique laboratory for the study of
the evolution of low mass stars.  The combination of their extreme
ages, compositions and dynamics also allows us a glimpse at the early
history of the Milky Way and the processes operating during its
formation. These aspects become even more significant in the context of
the star-to-star light element inhomogeneities found among red giants
in every cluster studied to date. The large differences in the surface
abundances of C, N, O, and often Na, Mg, and Al have defied a
comprehensive explanation in the three decades since their discovery.

Proposed origins of the inhomogeneities typically break down into two
scenarios: 1) As C, N, O, Na, Mg, and Al are related to proton capture
processes at CN and CNO-burning temperatures, material cycled through a
region in the upper layers of
the H-burning shell in evolving cluster giants may be
brought to the surface with accompanying changes in composition;
see \cite{sweigart79} and \cite{charbonnel1994} for an introduction
to the relevant theory of mixing.
There is ample observational evidence that deep mixing  takes place 
during the red
giant branch (RGB) ascent of metal-poor cluster stars (see the
reviews of Kraft 1994,
Pinsonneault 1997, \& Gratton, Sneden \& Carretta 2004 and 
references therein). 2) It has also become
apparent that at least some component of these abundance variations
must be in place before some cluster stars reach the giant branch.
Spectroscopic observations of main sequence turn-off stars in 47 Tuc
(beginning with Hesser 1978, see Cannon \etal\
1998, Briley \etal\ 2004a and references therein)
demonstrate this.  Our
own work with large samples of low luminosity
stars in M71 (Cohen 1999, Briley \& Cohen 2001,
Ram\'{\i}rez \& Cohen 2002) and in M13 (Briley, Cohen \& Stetson
2002, 2004b and Cohen \& Melendez 2004)
have shown variations in CN
and CH-band, O and Na line strengths 
among subgiants and main sequence stars consistent with patterns found among
the evolved giants of these clusters.  The low
mass cluster stars we observe are, however, 
incapable of both deep dredge-up and significant
CNO nucleosynthesis while on the main sequence.  Hence
the early cluster material must have been at least partially inhomogeneous in
these elements or some form of modification of these elements 
took place within the cluster. Suggested culprits include mass-loss
from intermediate mass asymptotic giant branch stars and supernovae
ejecta. \cite{cannon98} present an excellent discussion of these
possibilities.

Thus the observed light element inhomogeneities imply that there is
some aspect of the structure of the evolving cluster giants which
remains poorly understood (the deep mixing mechanism), that the early
proto-clusters may have been far less homogeneous, that intermediate
mass stars may have played a greater role in setting the composition of
the present day low mass stars than previously thought, etc.
It is this set  of issues that we explore in the present series of papers.
In our earlier work we studied the C and N abundances for large
samples of stars in the globular clusters M71 \citep{cohen99,briley01}, 
M5 \citep{cohen02} and M13 \citep{briley02,briley04b} (collectively
denoted GC--CN).  We consider here
the extremely metal poor globular cluster M15.
In this range of metallicity and luminosity (i.e. \teff), the CN bands
at 3880 and at 4220~\AA\ in the spectra of the M15 stars are
too weak to be useful, so we rely on
the CH and NH band strengths to derive C and N abundances
for a large sample of stars in M15.
We adopt values
from the on-line database of \cite{harris96}
for the apparent distance modulus of M15 at $V$ of 15.31 mag
with a reddening of E(B--V) = 0.09 mag, supported by analysis of deep 
HST photometry by \cite{recio04}.
We adopt the metallicity
[Fe/H] = $-$2.2 dex, also from \cite{harris96}; \cite{sneden91}
in a high dispersion abundance analysis of a large
sample of stars on the upper giant branch of M15 found a somewhat
lower value.

We describe the sample
in \S\ref{section_phot} and \ref{section_spec}.  We outline our
measurement of the molecular band indices and their interpretation
in \S\ref{section_indices}.  With an 
assumption about
the O abundance, these are converted into C and N abundances,
from which we find an anti-correlation between C and N in
\S\ref{section_cnabund}.  A consideration of the need for
ON burning follows in \S\ref{section_on}.
A discussion of our results together with a
comparison with the trends seen among the red giants in M15
with our earlier work, which now covers
four globular clusters spanning a wide range of metallicity,
combined with data from the literature, is
given in \S\ref{section_othergc}, while \S\ref{section_mix}
discusses the implications of our results for the mechanism producing
the C and N differences.  Inferences 
we can draw from this
for the formation and early chemical history of globular clusters
are given in \S\ref{section_chem_evol}.
A brief summary concludes the paper.

\section{Photometric Databases \label{section_phot}}

The optical photometry of M15 employed here was carried out as part of a larger
program to provide homogeneous photometry for star clusters and nearby
resolved galaxies \citep{stetson00}.  The general characteristics of the
photometric database are described in detail in
\cite{cohen02} for the case of M5; the case of M15 is similar.
At the time the present
photometry for M15 was derived, the corpus of M15
images in Stetson's database included some 307 images in $B$, 340
images in $V$, and 181 in $I$.  The images did not all cover the
same region of sky, of course, and any given star fell within no
more than 235 $B$ images, 239 $V$ images, or 179 $I$ images.
A network of local standards supported appropriate transformations for
those images taken under non-photometric conditions.
In our experience, photometry from datasets such as those employed here
typically display an external accuracy of order 0.02$\,$mag per
observation;
this level of observation-to-observation scatter is probably
dominated by temporal and spatial fluctuations in the instantaneous
atmospheric extinction, and probably also by the difficulty of obtaining
truly appropriate flat-field corrections in the presence of such effects
as scattered light, ghosts, fringing
and spectral mismatch between the flat-field illumination and the
astronomical scene. The fundamental system is that of \cite{landolt92}.

The absolute astrometry of our catalog is based upon the United States
Naval Observatory Guide Star Catalogue~I (A~V2.0; henceforth USNOGSC,
Monet \etal\ 1998),
access to which is obtained by PBS
through the services of the Canadian Astronomy
Data Centre.  Throughout the
region of our field that is well populated by USNOGSC stars (including
essentially all of the stars in our present spectroscopic sample), 
we expect
systematic errors of our right ascensions and declinations on the system
of the USNOGSC to be  $<0.1\,$arcsec.   Individual
{\it random\/} errors in our coordinate measurements are 
probably not much
better than 0.02$\,$arcsec on a star-by-star basis, the errors becoming
somewhat worse than this for the fainter and more crowded stars in our
photometric/astrometric sample.

The alignment images for the slitmasks used with the
Low Resolution Imaging Spectrometer (LRIS) \citep{lris_ref} 
were taken with $V$ or $I$ filters.  Although these exposures were
short (1 sec typically),
they were used to determine $V$ mags for the
faintest stars, particularly those that were somewhat crowded, 
in our sample in M15.

To broaden the wavelength range of our photometry, we attempted to obtain
infrared colors from the 2MASS database \citep{2mass1,2mass2} 
for the stars in our sample.  However,
many of them are too faint to be included therein.
Thus in Sep. 2004 we observed the fields of our M15 sample
with the Wide Field Infrared Camera \citep{wilson03}
at the 5-m Hale Telescope for the purpose of establishing 
reliable J,K magnitudes
for the fainter stars in our sample.  The 2MASS colors
of nearby isolated
somewhat brighter stars were used to calibrate our WIRC photometry.  
Total integrations of 10 min to 30 min for each of the two filters
in each of the two fields were obtained.  These images were reduced
using Figaro \citep{shortridge} and DAOPHOT \citep{stetson87}.

Most other recent photometric studies of M15 \citep{buonanno87,dacosta90}
do not reach as faint as the bulk of our sample.  The deep $B,V$ CMD
study of \cite{durrell92}, which focuses on the age of the cluster,
its distance, and its luminosity function along the main sequence,
does not cover the full sample of our stars.  

Stars are identified in this paper by a name derived from their J2000
coordinates, so that star C12345\_5432 has coordinates
21 12 34.5~~+12 54 32.

\section{Spectroscopic Observations \label{section_spec}}

The initial sample of stars consisted of those
from the photometric database located more than
180 arcsec from the center of M15 (to avoid crowding) with
$16.5<V<18.5$ and with $B-V$ within 0.06 mag of the cluster
locus, which we take as $B-V = 0.73 - 0.062(V-16.5)$ mag.
The main sequence turnoff of M15 is at $V \sim 19.2$ mag,
so these stars include subgiants as well as low luminosity giants near the
base of the RGB.
(A preliminary
version of the photometric catalog described in \S\ref{section_phot}
was used for this purpose.)
From this list, two slitmasks containing about 25 slitlets
each were designed using JGC's software.  The center of the first
field was roughly 3.2 arcmin E and 0.7 arcmin N of the center
of M15, while the center of the second field was located roughly
2.8 arcmin W and 2.9 arcmin S of the cluster center.

These slitmasks were used with LRIS
at the Keck Observatory
in June 2003.  Three 1200 sec exposures were obtained for the second mask
and two 1200 sec exposures for the first slitmask.  The airmass was
less than 1.06 for all exposures, so differential refraction did not
play a role even at 3200~\AA.
The exposures were dithered by moving
the stars along the length of the slitlets by 2 arcsec 
between each exposure.  Because of the crowded fields,
there was often more than one suitably bright object in each
slitlet.  
The width of the slitlets was 0.8 arcsec,
narrower than normal to enhance the spectral resolution.
LRIS-B \citep{lris-b} 
was used with a 400 line grism giving a dispersion of
1.0\,\AA/pixel (4.0\,\AA\ resolution for a 0.8 arcsec wide slit).
This gave good coverage of the region from 3000 to 5000\,\AA,
including the key NH band at 3360\,\AA\ and the G band of CH
at 4300\,\AA.  The CN bands at 3880 and 4200\,\AA, while included
within the spectral range covered, are for most of these low luminosity 
very metal-poor stars
too weak to be measured with precision; they were not used at all.
The red side of LRIS was configured to use a 1200 g/mm
grating centered at H$\alpha$ with the intention of providing
higher accuracy radial velocities.  The dispersion is then
0.64\,\AA/pixel (29 \kms/pixel) or 1.9\,\AA/spectral resolution element.
Figaro \citep{shortridge} scripts were used for the data reduction.

The original detector of LRIS-B
was upgraded to a new one with much higher UV 
sensitivity prior to these observations.  This was crucial to the
success of our effort.  However, there were some unexpectedly
severe reflection problems in our blue-channel spectra.  These
were perhaps exacerbated by the many bright stars in the field.
The reflections were non-dispersed, aligned along the slit,
and several times the height of a stellar image.  They were
removed partially by sky subtraction, but the resulting spectra had
to be hand checked, with additional corrections applied as necessary.
This was done for each individual exposure, then the resulting spectra
for each star were summed.   In addition, the LRIS-B images
were not flattened, as there is no suitably bright 
featureless UV calibration lamp at the Keck Observatory.  The pixel-to-pixel
variation is small in these detectors, and each spectrum is the
sum of several exposures which fell in different locations
on the detector array.  Furthermore, the projected image size
of a  point source along the slit has a FWHM of 3 to 4 pixels,
so many pixels are sampled in forming the final 1D spectrum for each star.  
We therefore are confident that the lack of flat fielding does 
not introduce spurious small scale features.  

The final spectrum summed over the three 1200 sec exposures of a M15 star
in our sample 
with $V = 18.0$ mag has roughly 9000 detected
electrons per spectral pixel (1.0~\AA) in the region of the G band of
CH, and roughly 1100 detected electrons per spectral pixel in the
blue continuum bandpass for the NH feature. 
For a $V = 18.0$ mag star in
the slitmask with only two 1200 sec exposures (stars numbered C30...), 
there are roughly 6000 detected 
electrons per spectral pixel (1.0~\AA) in the region of the G band of
CH, and roughly 500 detected electrons per spectral pixel in the
blue continuum bandpass for the NH feature.  Thus the 
uncertainties in the measured CH and NH indices are not dominated by
Poisson statistics, even for the faintest stars in our sample in M15,
but rather by the various instrumental issues described above.

In addition to the primary sample described above, since
these fields are rather crowded, other stars sometimes serendipitously
fell into the slitlets.  If they were bright enough, their spectra 
were also reduced.  We refer to the latter as the secondary
sample.  As might be expected from the luminosity
function, most of the secondary sample consists of stars at
or just below  the
main sequence turnoff.

\subsection{Membership}

The galactic latitude of M15 is only $-27.3^{\circ}$.  Even though
our fields are as close to the center of the cluster as possible,
the cluster is more distant than ones we previously studied, 
and our sample consists of faint stars.
Establishing membership is a concern.
The primary basis for determining membership is through the spectra.
M15 is a very metal poor cluster, so that these moderate resolution
spectra are in themselves capable of confirming membership.
Table~\ref{table_phot} gives the 
$V, I, J, K$ photometry for the 68
sample stars we believe to be members of M15.  
Three non-members were culled from our sample as their spectra show
absorption features much stronger than expected; they are listed
at the end of this table.

\begin{figure}
\epsscale{1.0}
\plotone{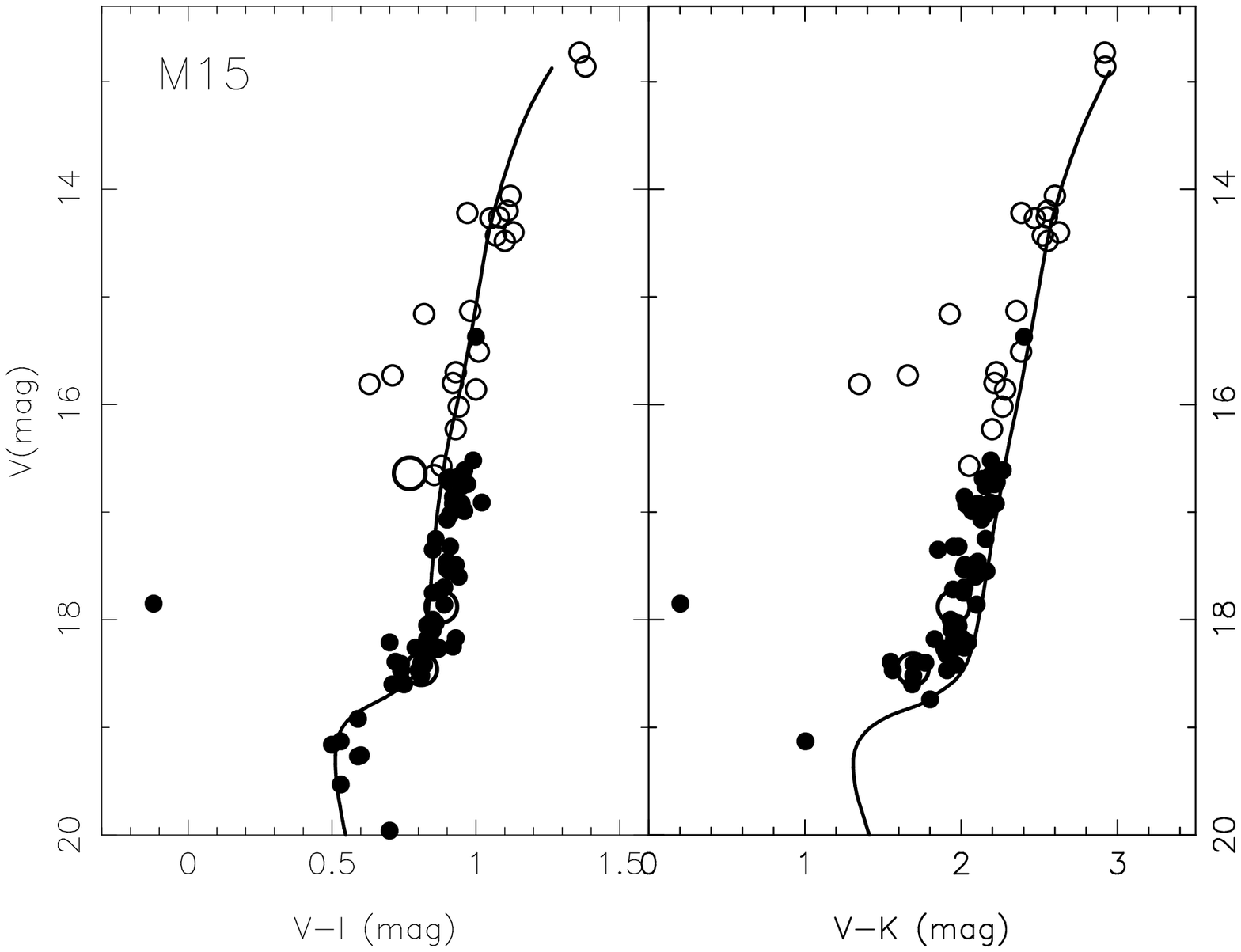}
\caption[]{The $V,I$ and $V,K$ CMDs for M15 are shown.  The sample stars
are indicated by small filled circles.    The large open
circles denote
spectroscopic non-members. Smaller open circles denote the members
of a Keck/HIRES sample to be discussed in Cohen \& Melendez (2005, in preparation). 
A 12 Gyr, [Fe/H] $-2.2$ dex isochrone
from \cite{yi01} is also shown shifted to our adopted cluster
distance and reddening.  
\label{fig_cmd}}
\end{figure}

The CMD diagrams can also be used to eliminate non-members.  
$B,V$ colors were used to select
the primary sample, but they (and other colors)
can provide constraints for the secondary stars.
Fig.~\ref{fig_cmd} shows the $V,I$ and $V,K$ CMD diagrams for our sample in M15
with a 12 Gyr isochrone with [Fe/H] $-2.3$ dex from \cite{yi01} superposed
in each case. 
The star at the lower left is a hot horizontal branch star; it is
part of the secondary sample.
The spectroscopic
non-members are indicated by large open circles; 
they all line on the cluster isochrone in the $V-K$ CMD, and
two do for the $V-I$ CMD, with the third only slightly off it.
Thus photometry alone,
while it can eliminate many non-members of M15 from our sample selected for
spectroscopy, is not sufficient in itself.

\begin{figure}
\epsscale{1.0}
\plotone{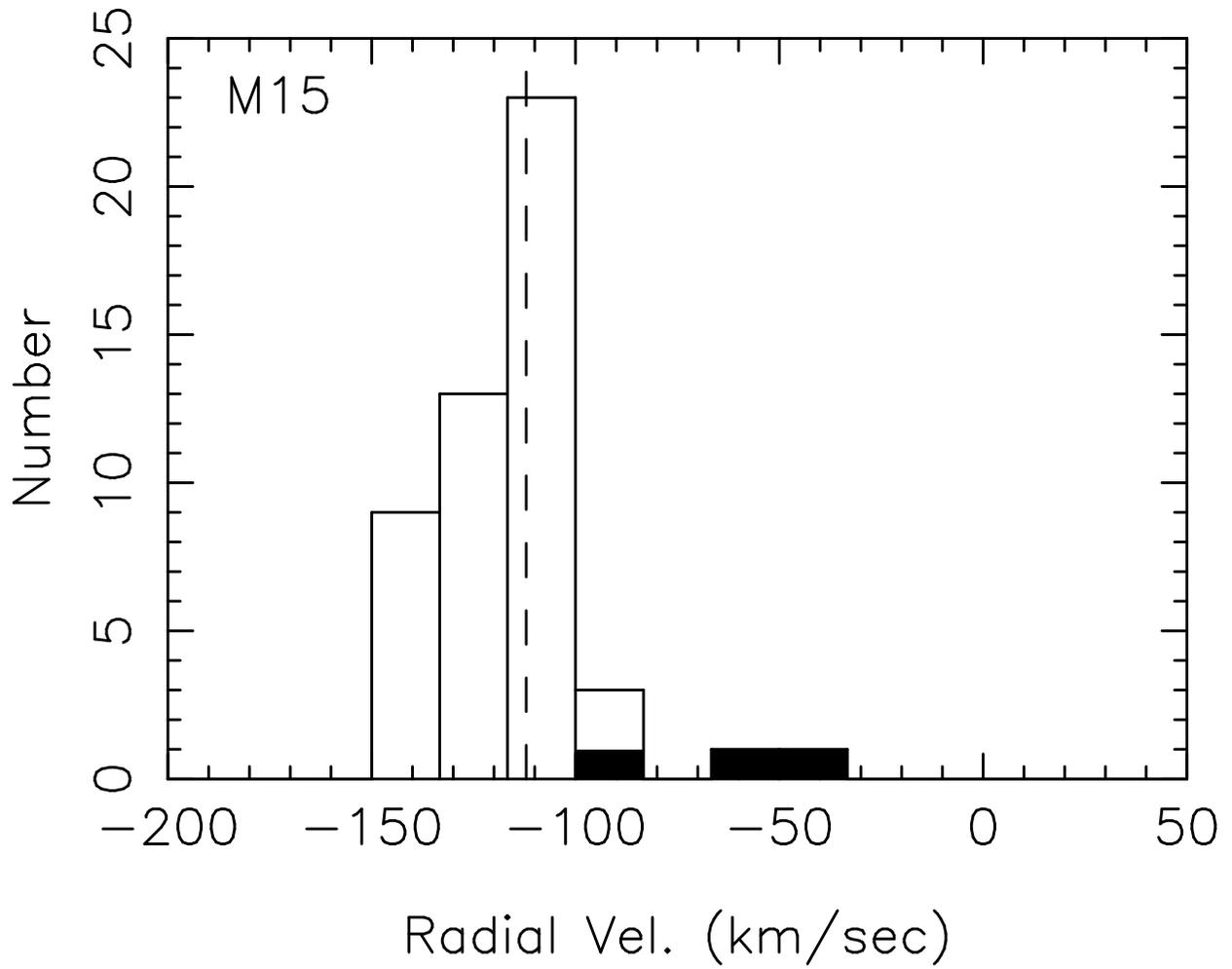}
\caption[]{The histogram of the radial velocities for 50
stars in our sample in M15 is shown.   The filled
area denotes the three stars non-members based on their spectra.
The dashed vertical line indicates the systemic velocity of the cluster.
\label{fig_vr}}
\end{figure}

The magnitude of the radial velocity of M15 is sufficiently high to establish
membership through measurements on the red spectra, which are of higher
spectral resolution than those from LRIS-B with the adopted
instrument configuration.   Given
the extreme metal deficiency of the stars and the foreground reddening,
which produces easily detectable interstellar NaD lines, we 
rely exclusively on
H$\alpha$ for this purpose.  Thus these $v_r$ measurements are not
of high accuracy, with typical uncertainty of $\pm30$~\kms.  A histogram 
of the radial velocities
for 50 stars from our sample is shown in Fig.~\ref{fig_vr}.  The three
spectroscopic non-members culled from our sample are indicated.
This figure demonstrates
that the vast majority of the stars in our sample are members of M15.

The spectra of star 
C29445\_0952 (V=16.88) in our sample are more 
extended along the slit than that of a point source
and clearly
indicate that there are two different stars contributing.
The photometric database and the LRIS 
alignment images were checked; this object turns out to be a close pair
of separation 0.8 arcsec with a brightness difference of 2 mag. It was not
possible to separate the contributions of each to the spectra, so
data for this
object is included in the tables, but it is not shown in any of the figures.

\begin{figure}
\epsscale{1.0}
\plotone{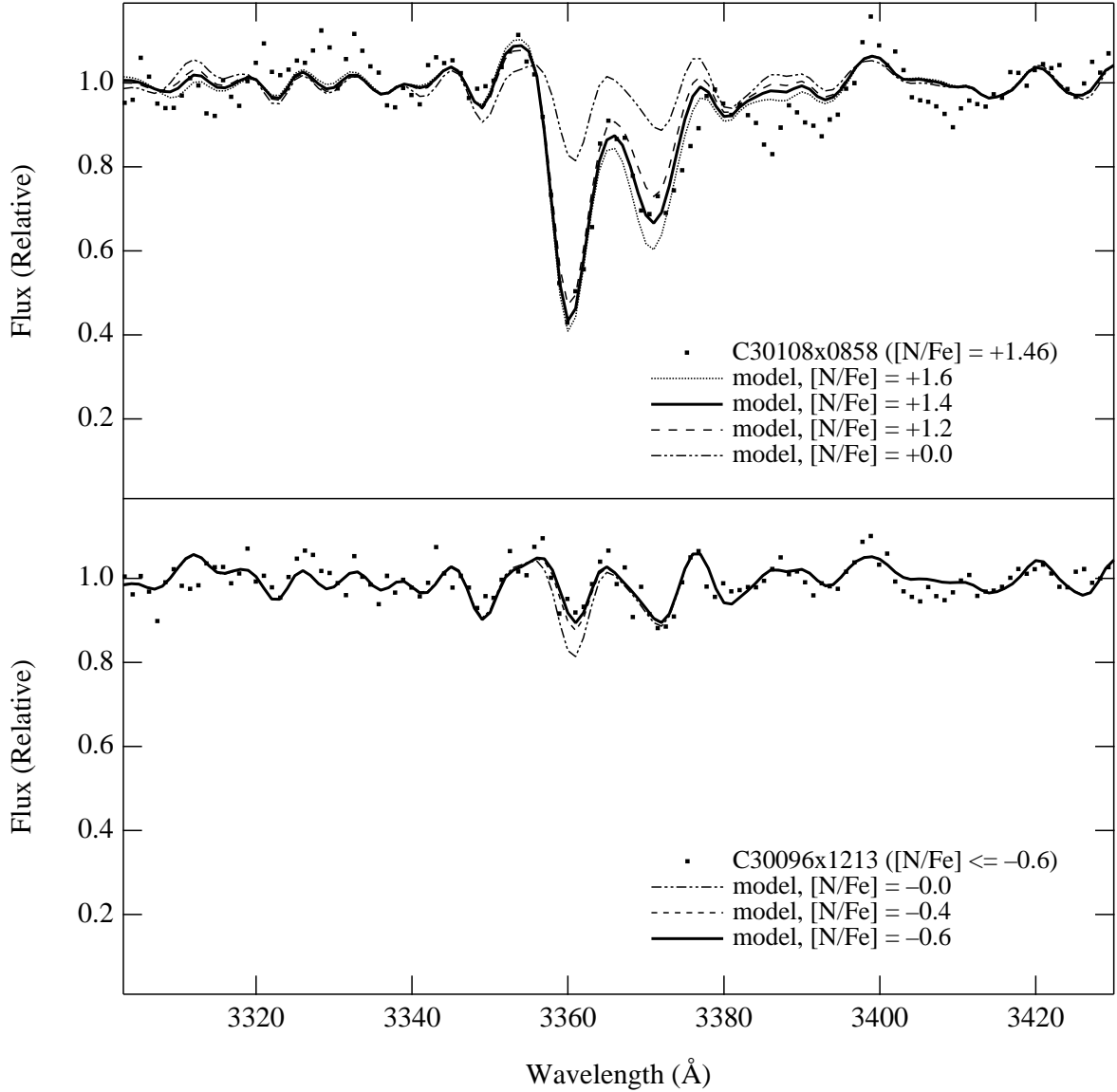}
\caption[jcohen_fig3.ps]{The LRIS-B spectra of two
members of M15 in the region of the NH band are shown (the points).
The stars are essentially identical in $V$ mag (V=16.92) and $V-I$ colors and
are both located at the base of the RGB branch.
Synthetic spectra with various values of [N/Fe] are superposed.
\label{fig_2spec}}
\end{figure}

Figure~\ref{fig_2spec} shows the region of the 3360\,\AA\ band of NH
in the spectra of two of the stars in the primary sample in M15.  These
stars have essentially the same stellar parameters (\teff\ $\sim5200$K and 
\grav\ $\sim$2.8 dex)
lying at about the same place in the cluster CMD, yet their
NH bands differ strongly. 
From this figure alone, we can anticipate one of the key results
of our work, the large scatter in C and N abundance we will find among
M15 members at the base of the red giant branch (RGB) at $17 < V < 18.5$ mag.

\section{Measurement of CH and CN Indices \label{section_indices}}

For each spectrum, indices sensitive to absorption by the
4300\,\AA\ CH band were measured as described in \cite{briley01}.
While for the NH band we could have used the double-sided
index described in \cite{briley93} and used by us for our small sample 
of spectra of M13 subgiants described in  \cite{briley04b},
we were concerned with the decline of the apparent continuum
level towards bluer wavelengths in the UV for
these LRIS-B/Keck spectra.  This is
presumably due to the wavelength dependence of both the stellar  flux
and the instrumental efficiency. 
It is extremely difficult to flux spectra taken through
slitmasks because of the varying slit losses and the possibility
of atmospheric dispersion affecting the spectra, although the latter
was, as discussed earlier, not a concern here.  Carrying
out the observations with the length of the slit set to
the parallactic angle, which is the usual method for
eliminating atmospheric dispersion for single slit observations,
cannot be used for multislit observations as the position
angle is fixed by the design of the slitmask.  It is for these
reasons that no attempt was made to flux the spectra.

Since the stars are so metal
poor, there are relatively few detected features in the region
between 3300 and 3430\,\AA\ besides
the desired NH band.  Thus we decided to normalize the stellar continuum
in the spectrum of each star, then find the absorption
within the NH feature bandpass.  
The continuum
fitting approach adopted here allows for a direct comparison between indices
measured from  the spectra and those computed from theoretical spectral 
synthesis without the need for any slope corrections.
The feature bandpass for NH adopted here was 3354 to 3375\,\AA\ (in the
rest frame of M15).  This was done by
fitting a second order polynomial to the bandpass 3300 to 3430\,\AA,
masking out the region of the NH band.  The polynomial fitting used
a 6$\sigma$ high and 3$\sigma$ low clipping, running over a 5 pixel
average.  There was little change in the measured NH indices as compared
to simply applying the two-sided continuum and feature bandpasses.
This reflects our educated guess for choice of the continuum
bandpasses, which are more or less symmetrically 
distributed about the center of the feature bandpass.  

\begin{figure}
\epsscale{0.9}
\plotone{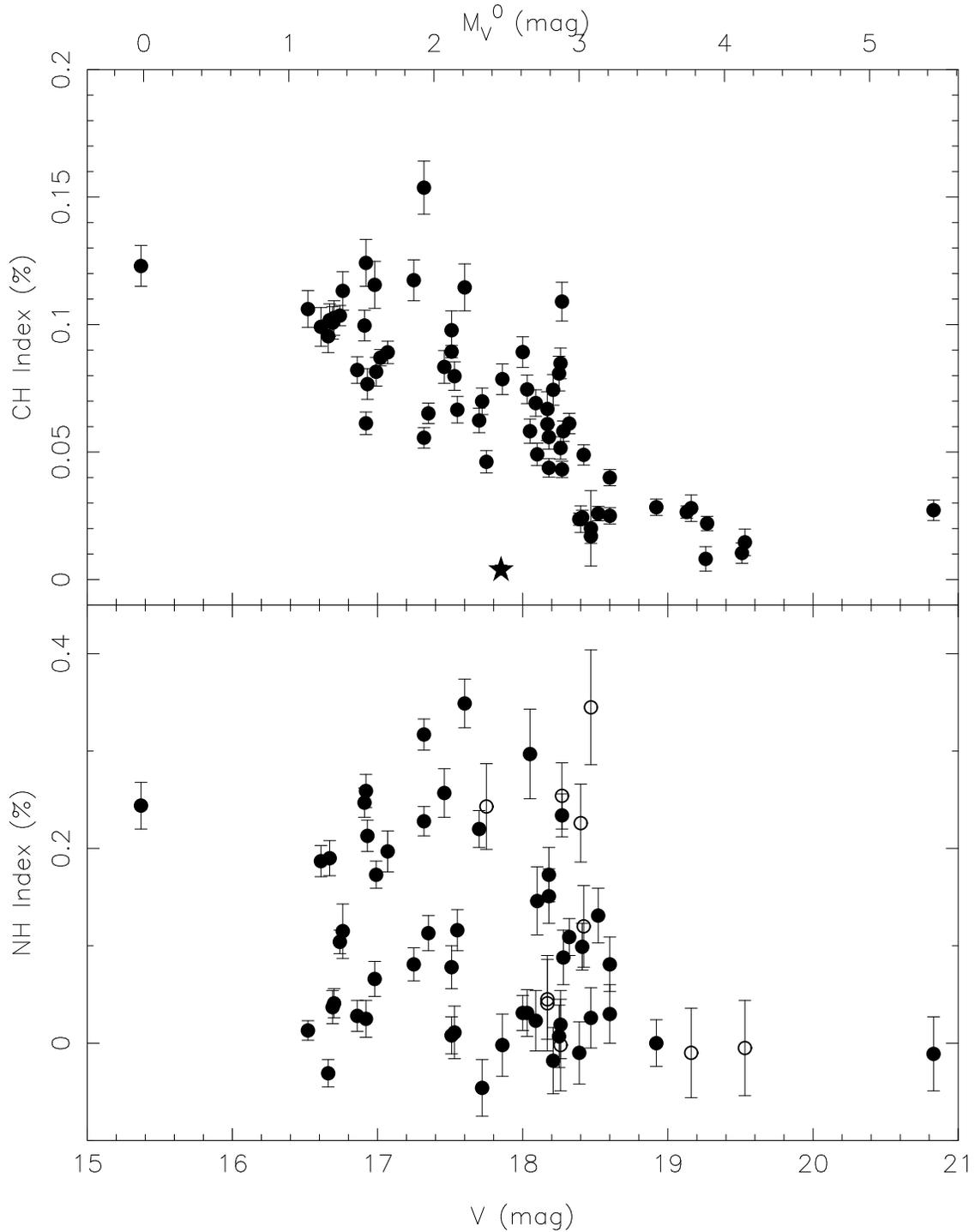}
\caption[jcohen_fig4.eps]{
The measured I(CH) and I(NH) indices are plotted for the program 
stars as
a function of $V$; the top axis is $M_V^0$. The open circles in the 
I(NH) (lower) panel denote
the stars with the smallest signal level at 3350~\AA. Large and 
significant star-to-star
differences exist in both CH and NH band strengths among these stars.
The decreasing spread in indices with luminosity is the result of 
increasing
temperatures near the main sequence turn off.
\label{fig_chcn_indices}}
\end{figure}

The measured indices are listed in Table~\ref{table_obs_inds} and plotted 
in Fig.~\ref{fig_chcn_indices}.  The upper axis of these figures
is $M_V^0$, and the RGB bump in M15 occurs at $V=15.41\pm0.04$ mag
\citep{zoccali99}.  Approaching the MSTO, \teff\ suddenly increases, and the
molecular bands we study here become much weaker; they are effectively
undetectable in the present spectra.  This produces the sharp drop
in measured indices apparent at $V \sim18.5$ in Fig.~\ref{fig_chcn_indices}.
It is not until $\sim$2.5 mag below the MSTO that \teff\ as cool as 5600~K
is again reached, at which point molecular bands can again be 
expected to be detectable
in M15 main sequence stars with similar quality spectra to those presented here
for its subgiants.

The error bars given in Table~\ref{table_obs_inds}
(drawn in Fig.~\ref{fig_chcn_indices} at 2$\sigma$) have been calculated strictly 
from Poisson
statistics based on the signal present in the feature and continuum 
bandpasses.  The CH index, in particular, is very
robust; independent measurements by each of the first two authors
using both a double-sided index and continuum fitting as described above
always agreed to within 2\% for each star in the sample 
in M15.  Every star in the sample has a measurement of I(CH).
I(CH) is positive for all the stars in the sample, and 
exceeds 1\% for all stars except C29424\_0729
on the extended BHB and one star near the main sequence
turnoff at $V = 19.26$ mag.  We do not expect to see any CH
for the extreme BHB star nor for
the stars just at the turnoff.  The measured
I(CH) values for them range from 0 to 4\%, with most
having I(CH) $\le$2\%.
These include the faintest stars in the sample, and their I(CH) measurements
appear valid, again suggesting that the
measurements of the CH indices at all luminosities considered here are
robust.
As can be seen in Figure \ref{fig_chcn_indices}, substantial star-to-star
differences occur among the I(CH) indices for the stars in our M15 sample,
even if one considers only  stars in similar evolutionary states.

Obtaining high quality spectra in the region of the NH band
at 3360~\AA\ for such faint stars is not easy.
There are five stars with poor quality spectra in the NH region
for which no measurement of the NH index was attempted.  These arose
as a result of low signal level, the location of the spectra
with respect to the edges of the slit (a problem more common
among the secondary stars), or a bright reflection falling on or
very near the NH feature. 
There are 10 additional stars with low signal level,
but measured I(NH), which are marked by open circles instead
of filled circles in the figures; only one of  these is brighter
than $V=18.1$.  Among the stars brighter than $V=18.3$
with measured $I(NH)$, there are five with negative $I(NH)$.
Of these, three have I(NH) $ \ge -2$\%\, while the
smallest is $-5$\%.   Fig.~\ref{fig_chcn_indices} demonstrates
that a very substantial range exists in I(NH) among stars
of similar evolutionary state for $17 < V < 18~(1.6 < M_V^0 < 2.6)$ mag.

In examining Fig.~\ref{fig_chcn_indices}, aside from the extreme BHB star,
there are no obvious anomalous stars in our sample in M15.
Furthermore, there is no obvious bimodality among either the
CH or NH indices.
Over a small range in \teff, there appears to be 
an anti-correlation
of the strength of the CH and NH bands.  The correlation is not
perfect, but given the weakness of the molecular features in these
low luminosity stars in  M15, is reasonably convincing.  We return
to this issue in \S\ref{section_abund} once C and N abundances are derived
for the stars in our M15 sample.

The large range in C abundances which we suspect to be present in the
M15 subgiant sample creates an unusual situation with regard to the
expected strength of the CN features.  Normally, since there is
more carbon than nitrogen, the N abundance controls the amount
of CN.  However, if C is highly depleted, there can be fewer carbon
atoms per unit volume than nitrogen atoms, and C will control the
formation of CN, as suggested by Langer (1985).  
Since we are using the NH band to deduce N abundances, this is not
an issue here.

\section{Comparisons with Synthetic Spectra \label{section_cnabund}}

Clearly the pattern of abundances underlying the CH and CN band
indices of Fig.~\ref{fig_chcn_indices} cannot be interpreted on the
basis of band strengths alone - we must turn to models.
The technique employed is similar to that of \cite{briley01}, where the
region of the CMD of interest is fit by a series of models whose 
parameters are taken 
from (in the present case) a Y2 \citep{yale04} 12 Gyr isochrone
with Z = 0.000400, Y = 0.230800, and [$\alpha$/Fe] = 0.6.
The set of representative model points
are listed in Table \ref{table_modelch}.  Model stellar
atmospheres were then generated using
the Marcs model atmosphere program \citep{marc75} at the
\teff, \grav\ of these points.  Logarithmic 
solar abundances of Fe, C, N, and O are assumed to be  
7.52, 8.62, 8.00, and 8.86 dex, respectively, on a scale of H = 12.
These are somewhat higher than the latest solar abundances
inferred from 3D hydrodynamic models by \cite{asplund05}, but we
continue to use them for consistency with our previously published
papers in this series. 
Furthermore, our molecular transition probabilities were scaled to  
fit the solar spectrum with our adopted solar abundances when they  
were first adopted (e.g., \cite{bell94}).
From each model, synthetic spectra were calculated using the SSG 
program (Bell \&
Gustafsson 1978; Gustafsson \& Bell 1979; Bell \& Gustafsson 1989; 
Bell, Paltoglou,
\& Tripicco 1994) and the line list of \cite{trip95}.
The spectra were calculated from 3,200 to 5,500\,\AA\ in 0.05\,\AA\ steps
with a microturbulent velocity of 2 km/s, an [O/Fe] abundance of +0.20 dex,
\ciso = 10, but with differing C and N abundances.
The resulting spectra were then smoothed to the resolution of the observed 
spectra and the corresponding I(CH) and I(NH) indices measured. 
By construction, no zero point shifts were necessary.

\begin{figure}
\epsscale{0.7}
\plotone{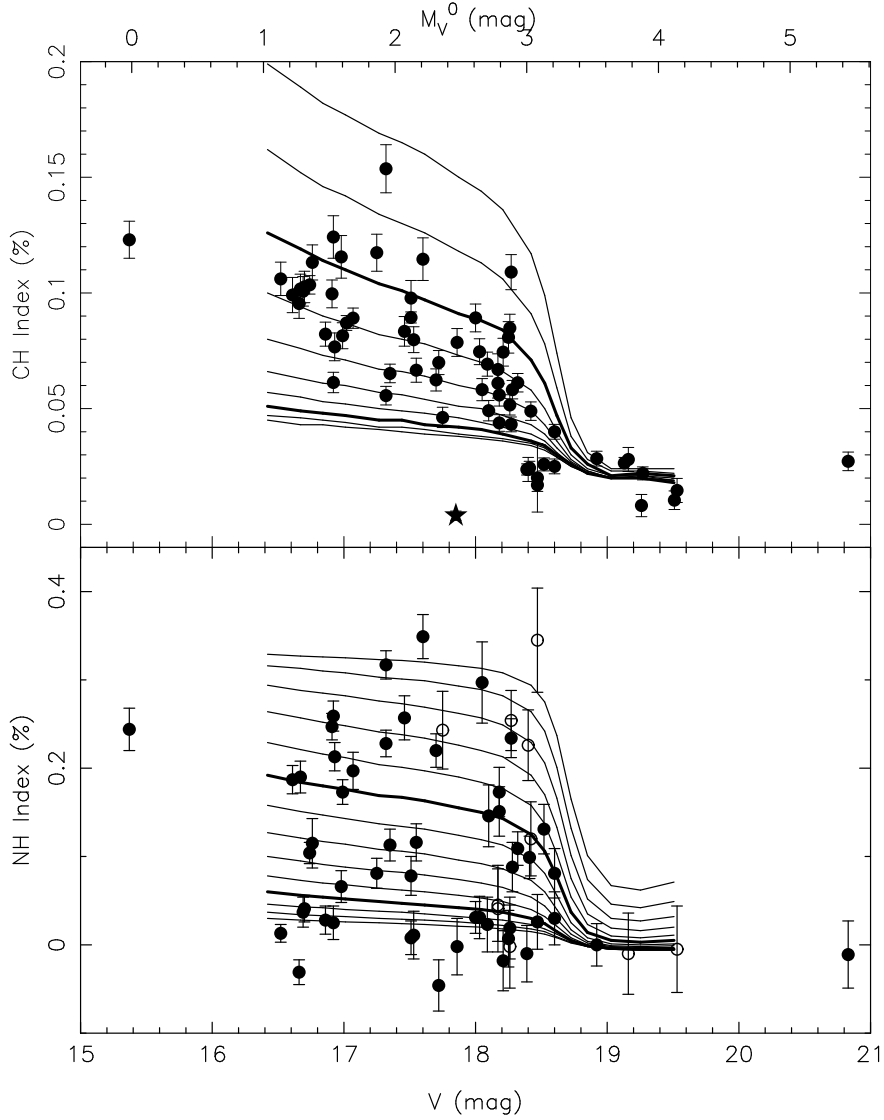}
\caption[]{
The measured I(CH) and I(NH) indices are plotted for the program 
stars as
a function of $V$; the top axis is $M_V^0$. The model I(CH) and I(NH)
are superposed, with thick lines denoting [C/Fe] = 0.0
and $-1.0$ dex in the upper panel, and [N/Fe] = 
0 and +1.0 dex in the lower panel. The open circles in the lower panel denote
the stars with the smallest signal level at 3350~\AA. Large and 
significant star-to-star
differences exist in both CH and NH band strengths among these stars.
The decreasing spread in indices with luminosity is the result of 
increasing
temperatures near the main sequence turn off.
\label{fig_chcn+model}}
\end{figure}

The values of I(CH) from 18 sets of models with [C/Fe] from 
$-1.4$ to +0.4 dex in steps of 0.2 dex are plotted with the observed indices in
Figure~\ref{fig_chcn+model}.  
I(NH) was computed from the same set of models for 
[N/Fe] from $-0.6$ to +2.0 dex in steps of 0.2 dex.  The corresponding
I(NH) values are also shown in this figure.
The spread in [C/Fe] and [N/Fe] among the M15 SGB stars appears well 
represented by this range.
Clearly the star-to-star spread in N abundances of the
stars at the base of the RGB in M15 approaches a factor of 10. 

The resulting I(CH) and I(NH) indices predicted via synthetic
spectra from the grid of model atmospheres are listed in 
Tables~\ref{table_modelch} and \ref{table_modelnh}.

\section{Inferred C and N Abundances Among the Subgiants 
\label{section_abund}}

\begin{figure}
\epsscale{0.9}
\plotone{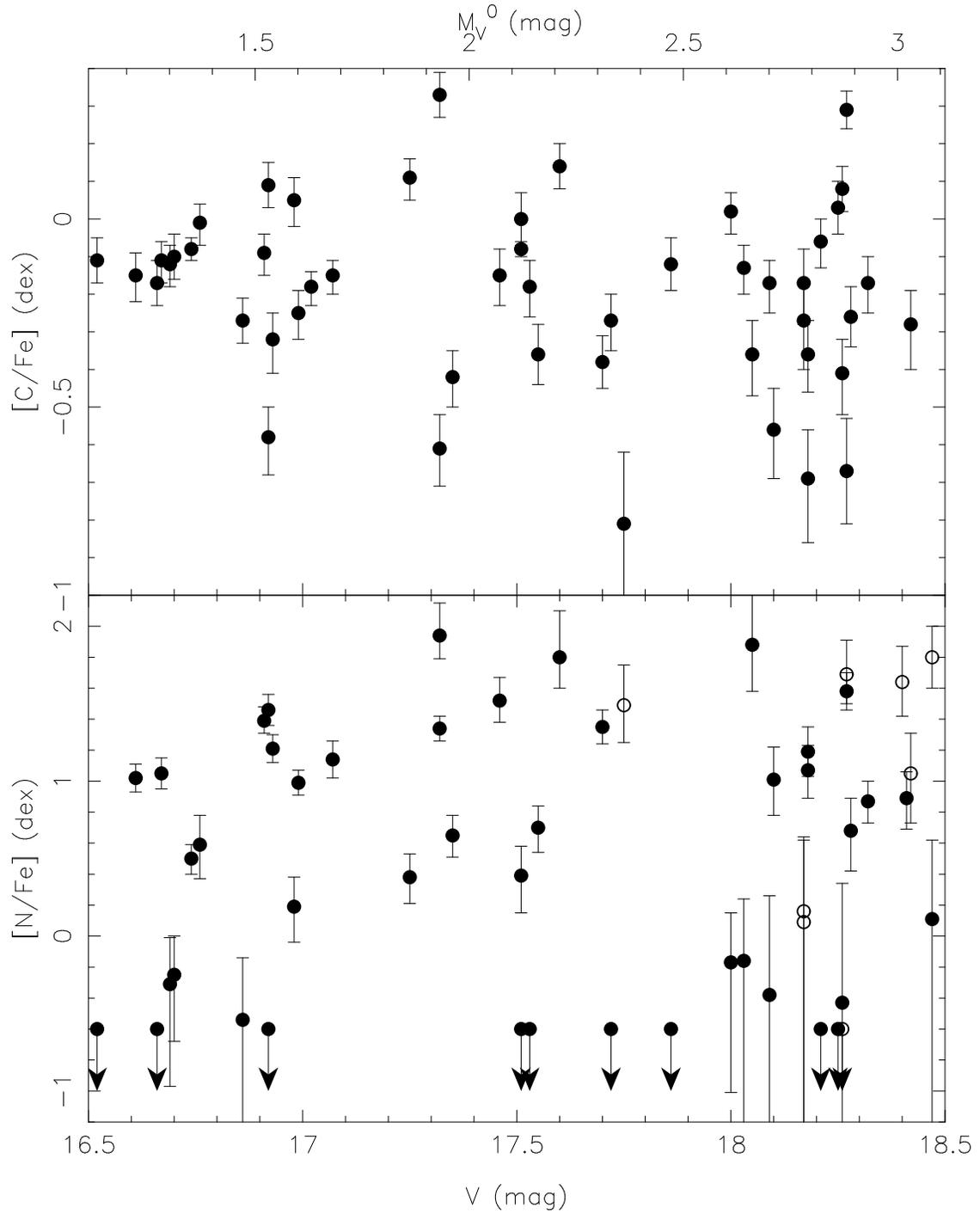}
\caption[]{The C and N abundances with respect to Fe are 
plotted for the program 
stars in M15 as
a function of $V$; the top axis is $M_V^0$. The open circles in the
[N/Fe] (upper) panel denote
the stars with the smallest signal level at 3350~\AA. Large and 
significant star-to-star
differences exist in both CH and NH band strengths among these stars.
The decreasing spread in indices with luminosity is the result of 
increasing
temperatures near the main sequence turn off.
\label{fig_cn_abund}}
\end{figure}

To disentangle the underlying C and N abundances from the
CH and NH band strengths, we have fit the [C/Fe] and [N/Fe]
abundances corresponding to the observed I(CH) and I(NH)
indices of the SGB stars in M15.
Since we are using the NH band instead of a CN band,
the coupling between the assumed abundance of C and the
deduced abundance of N is minimal and we employ here the
same technique of Briley \etal\ 2002, 2004b (our M13 analysis):
the model isoabundance curves of Table~\ref{table_obs_inds}
were interpolated to the $M_V$ of each program star using cubic splines,
and the observed index converted into the corresponding
abundance based on the synthetic indices at that $M_V$.
The resulting C and N abundances are plotted in Figure~\ref{fig_cn_abund} 
and listed in Table~\ref{table_obs_inds}.

The error bars were determined by repeating the process
while including shifts in the
observed indices of twice the average error among the SGB
indices as derived from
Poisson statistics (0.005 in I(CH) and 0.02 in I(NH)).
The shifts were included in opposing directions (e.g., +0.005 in
I(CH) and --0.02 in I(NH),
followed by --0.005 in I(CH) and +0.02 in I(NH)) and reflect
likely errors in the
abundances due to noise in the spectra.

Plotted in Figure~\ref{fig_2spec} we show synthetic spectra 
corresponding
to two SGB stars (one NH strong, the other NH weak) and a variety of
N abundances.
Also shown are the abundances resulting from index matching, which agree
well with what one would obtain from visual matching between observed
and calculated spectra.

The sensitivity of the derived C and N abundances to our
assumptions was also evaluated.
We chose four representative stars near the base of the RGB and again repeated
the fitting of the CN and CH band strengths with different values 
of [Fe/H],
[O/Fe], \ciso, etc.
These results are presented in Table \ref{table_changes}, where it 
may be seen that
the sensitivity of the derived C and N abundances to the choice of 
model parameters
is remarkably small (well under 0.2 dex for reasonably chosen 
values), as would
be expected from these weak molecular features.
We have also plotted the C and N abundances of our M15 sample as both functions 
of $V$ (Fig~\ref{fig_cn_abund}) and
$V-I$ colors (not shown) to evaluate possible 
systematic effects
with luminosity and temperature;  none appear to be present.

C and N abundances were not calculated for the stars at or below
the MSTO, as all abundance sensitivity is lost due to the high \teff\ and
resulting weakness of the molecular bands.
For the six  stars 
in our sample with I(NH) $\le 1$\%, essentially all of which
have I(NH)+1$\sigma[{\rm{I(NH)}}] > 0$, we assign upper limits
to [N/Fe], which are shown in Fig.~\ref{fig_cn_abund}.  This figure
immediately confirms the very large range in C and N abundances from
star to star at similar evolutionary stages in M15 previously
deduced from the appearance of the I(CH) and 
I(NH) - V mag plots (Fig.~\ref{fig_chcn_indices}).

\begin{figure}
\epsscale{0.8}
\plotone{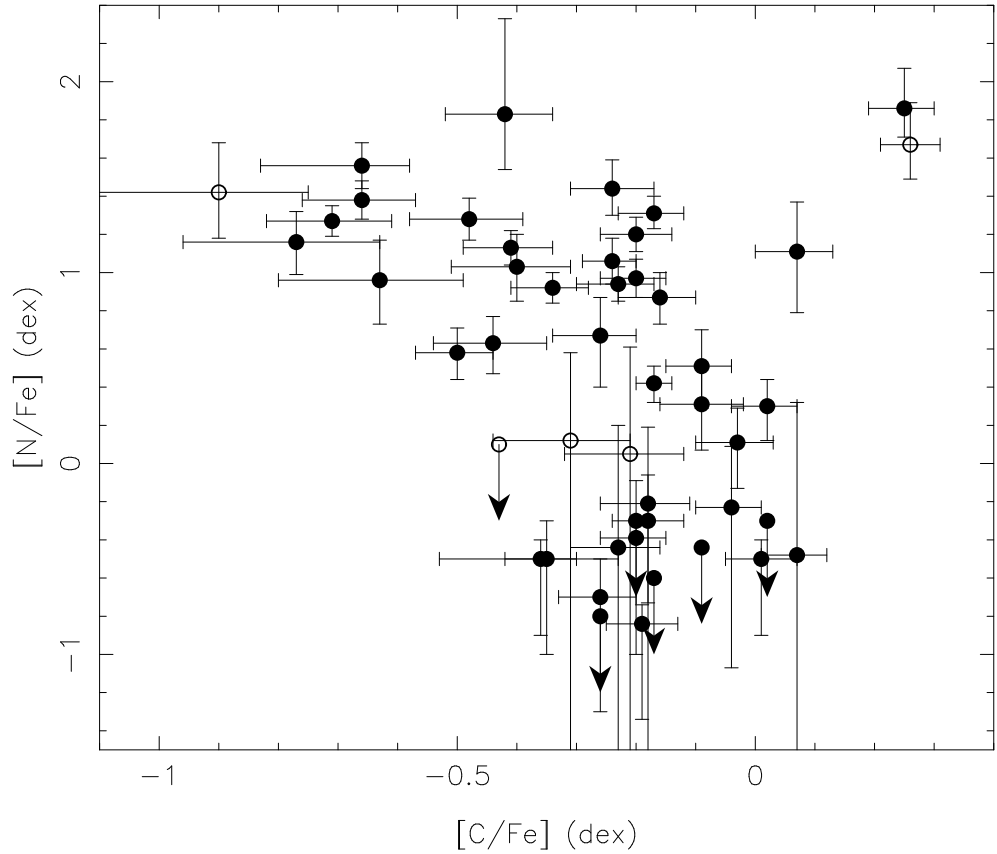}
\caption[]{
The derived [N/Fe]  for the M15 stars in Table
\ref{table_obs_inds} are plotted as a function of the [C/Fe]
abundances.
A C versus N anti-correlation is evident.
The presence of such an anti-correlation, although suggestive of the 
presence of
atmospheric material exposed to the CN-cycle, is difficult to
explain via internal processes given the evolutionary state of the
present sample of stars.
\label{fig_c_vs_n}}
\end{figure}

We next consider whether our data show any correlation between the 
C and N abundances we derive for M15 subgiants and lower RGB stars.
Fig.~\ref{fig_c_vs_n} shows a plot of [N/Fe] versus [C/Fe] for
the entire M15 sample.  An anti-correlation, with considerable
scatter, is apparent.  The scatter is consistent with the observational
errors, but there are a few outliers. 
In a sample of 70 objects with Gaussian errors, 
one outlier at the 2.5$\sigma$ level might be expected.
One of these, C30123\_1138 (V=18.27)
is among the stars with  low signal in the continuum near the NH band, hence I(NH)
with a high uncertainty, indicated 
in the figures by an open circle.  
The deviation of star
C29413\_1023 (V=17.32), with extremely enhanced C and N, from
the mean relation shown by the M15 sample in Fig.~\ref{fig_c_vs_n} is
of higher statistical significance.  This star will be discussed in Cohen \&
Melendez (2005), where additional relevant data will be presented.
We thus conclude, with the caveat of a very small number of
probable outliers, that an anti-correlation between
C and N is indeed found among the low luminosity sample of 68 stars in M15
studied here.

Evaluating the accuracy of our absolute abundance scale is more difficult as 
external comparisons
are limited. For the main sequence stars in 47 Tuc,
we can compare the results of Briley \etal\ (1991, 1994), carried out
in a manner fairly similar to the present work, with the independent
analysis of a different sample of stars by Cannon \etal\ (1998).
This suggests we may be systematically underestimating
the absolute C abundance by about 0.15 dex, and overestimating
the N abundance by about 0.2 dex. A second comparison 
is possible in M13.  We obtained C abundances using our procedures
as described here from newly obtained spectra
for previously studied bright giants in M13 precisely to address this
issue.  The mean differences in [C/Fe] for our results as compared
with literature
values  was 0.03$\pm0.14$ dex   for four stars
in common with \cite{smith96} and 0.14$\pm0.07$ dex for stars
also observed by \cite{suntzeff81} (if one extreme case is removed)
(see Briley \etal\ 2002 for details).  This is very reasonable
agreement.  It is clear that shifts in the absolute abundance
scale cannot
account for the large range in C and N abundances apparent in Figure
\ref{fig_c_vs_n}. We therefore conclude that the C versus N 
anti-correlation
among the low luminosity M15 stars in Figure \ref{fig_c_vs_n} is indeed real.

\subsection{From the RGB Tip to the Main Sequence Turnoff \label{section_trefzger} }

\begin{figure}
\epsscale{0.9}
\plotone{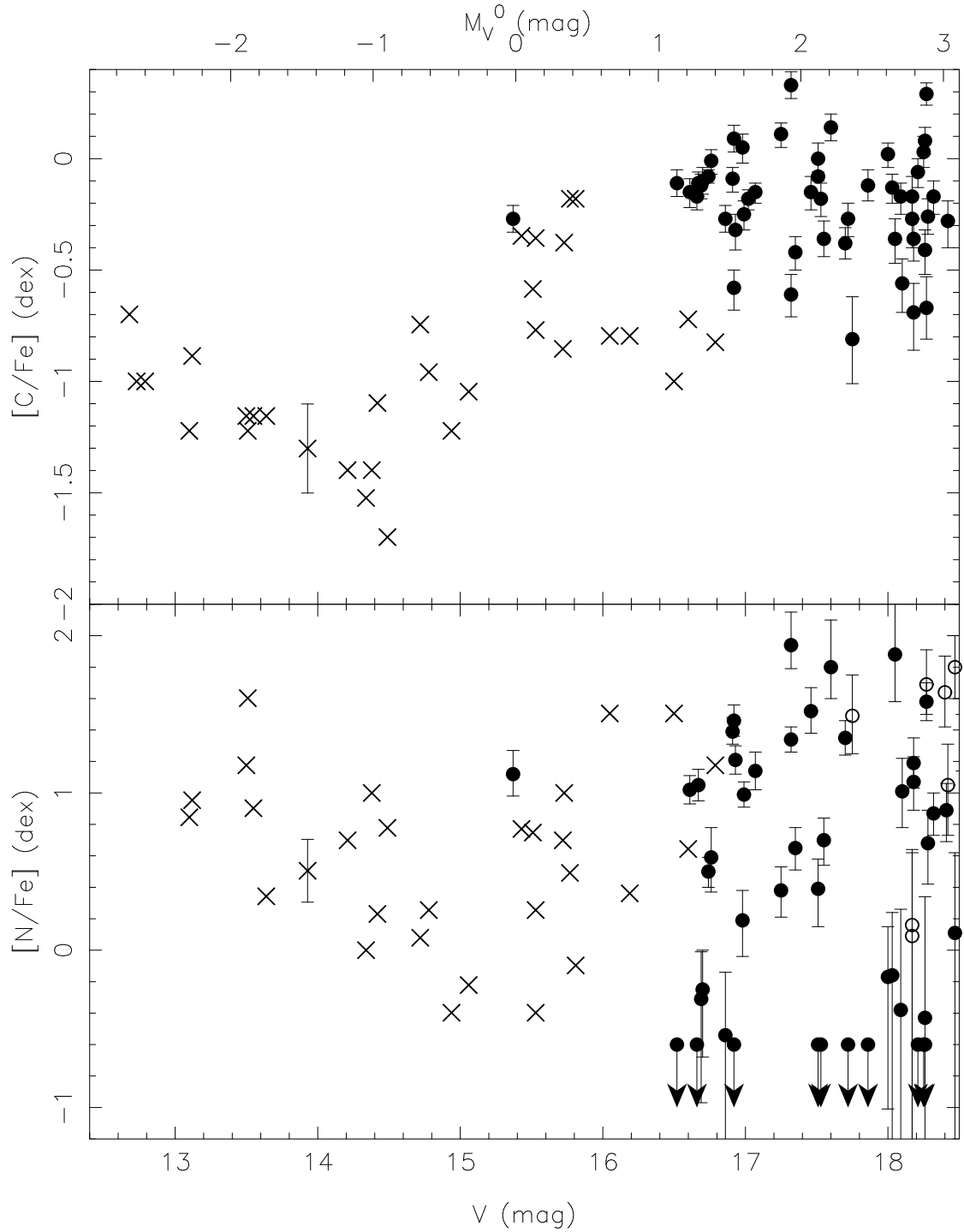}
\caption[]{The deduced [C/Fe] and [N/Fe] are shown as a function of $V$ mag
(with $M_V^0$ on the top axis) for the low luminosity stars
in M15 from our sample together with the C and N abundances
for luminous RGB stars from \cite{trefzger83}.  A typical uncertainty
for the latter is shown for a single star.
\label{fig_tref}}
\end{figure}

\cite{trefzger83} carried out an extensive analysis of C and N
abundances for the most luminous stars in M15.  We now combine
our results with theirs, thus sampling the [C/Fe] and [N/Fe] ratio
over the full range of luminosity from the RGB tip to the main
sequence turnoff in M15 in the two panels of Fig.~\ref{fig_tref}.

With the confidence that our
absolute abundance scale is reasonably secure and the hope
that the same holds for the work of \cite{trefzger83},
we assert that 
Fig.~\ref{fig_tref} shows a large range in [C/Fe] at low
luminosities, accompanied by a decrease in the mean [C/Fe] 
at about $V \sim$ 15 mag, which is essentially the location
of the RGB bump in this globular cluster.  We take this
as evidence of two separate mechanisms contributing to
the spread in the abundance of C and N in globular clusters.
At high luminosities near the RGB tip, 
we see evidence of the  
first dredge up, as expected from normal stellar evolution, plus the  
extra-mixing common among metal-poor cluster giants, with a decline  
in the mean C abundance of about a factor of 5 (0.8$\pm0.3$ dex);
the large uncertainty reflects the possibility 
that the absolute abundance scale of \cite{trefzger83} is
different from ours, a matter we plan to investigate in the
near future.  In metal-poor field giants 
\citep{gratton00,spite04} a similar drop of about a factor of 2.5
is seen in the C abundance at about the luminosity of the RGB bump.

These studies of field giants show an increase in N abundance of about a factor of
4 at ${\sim}L$(RGB bump), a drop in the Li abundance, and a decrease
in the $^{12}$C/$^{13}$C ratio as well.  There is some suggestion
of an increase in the mean N abundance for stars in M15 more luminous
than $L$(RGB bump) (Fig.~\ref{fig_tref}), but it is less clear cut
than the drop in mean [C/Fe] there.  There is also a potential concern of 
bias, in that \cite{trefzger83} could not reliably detect NH bands
weaker than those included here; their paper contains several
non-detections which were not plotted in Fig.~\ref{fig_tref}.

This is the same phenomenon we identified 
earlier in M13 \citep{briley02,briley04b},
where the mean C/Fe and spread about that value were constant from
the subgiants to below the main sequence turn off, but stars
near the RGB tip showed lower surface C abundance.  In that case,
due to the limited data from the literature for the luminous
RGB stars, we could not identify the luminosity at which the
transition occurred.  In M15, as is shown in Fig.~\ref{fig_tref},
that transition luminosity is reasonably well defined, and it is
$L$(RGB bump)

\section{ON Burning \label{section_on} }

\begin{figure}
\epsscale{1.0}
\plotone{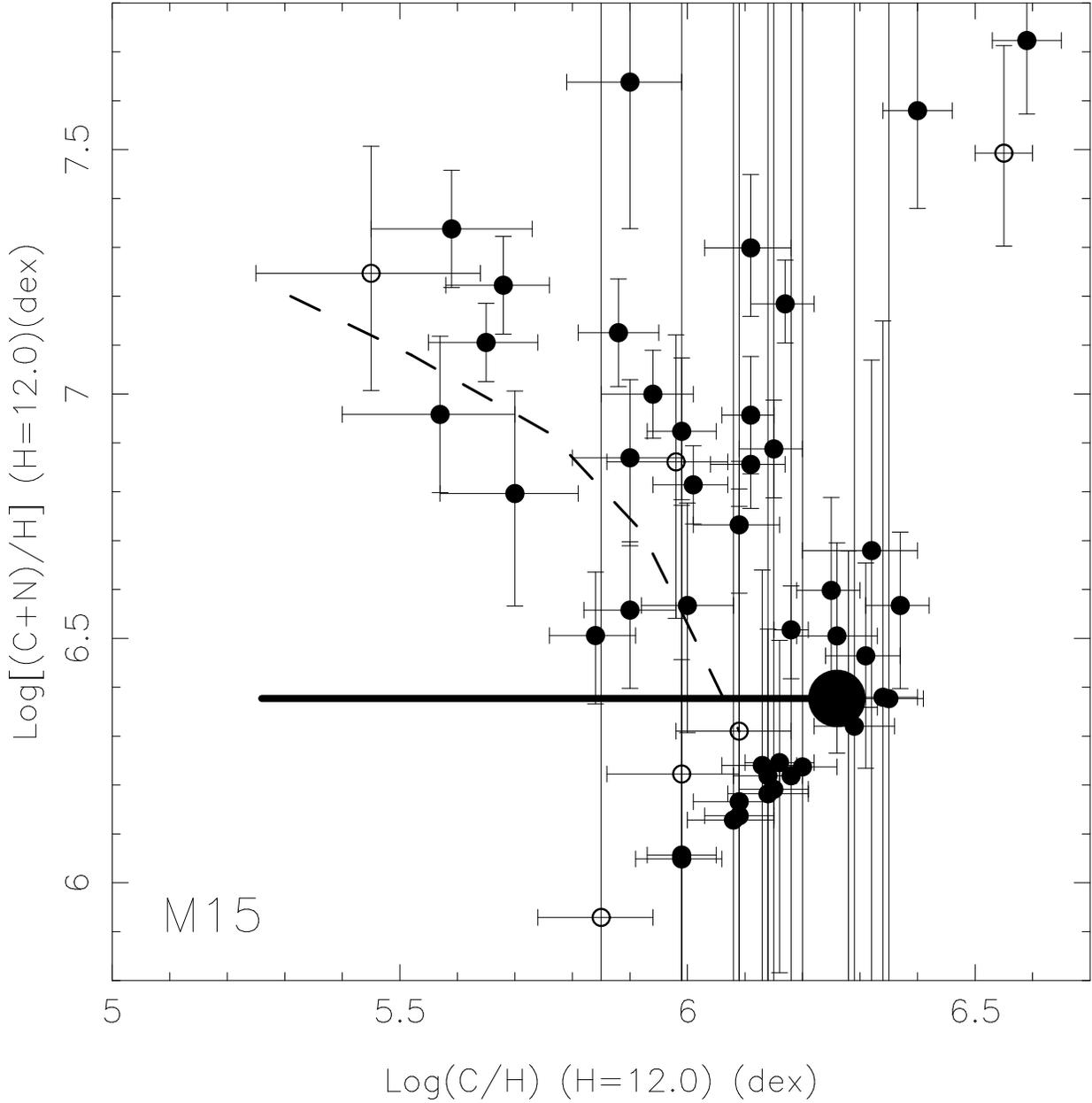}
\caption[]{The sum of the derived C and N
abundances is plotted as a function of the C abundance.  The large
filled circle marks the location for both C and N depleted by a factor
of 16, adopting the abundance of M15 of [Fe/H] = $-2.3$ dex, with
C/N at the Solar ratio.  The horizontal line extending to the left
of that represents the locus of points for C gradually being converted
into N, with the left end of the line having C/C$_0$ = 0.1.
The dashed line indicates the relationship, shown over its
full range, that prevails in M5
from our earlier work \citep{cohen02}.
\label{fig_sum_cn_m5}}
\end{figure}

We next examine whether converting C to N, presumably via the CN cycle,
is sufficient to reproduce the behavior we have found for our M15 sample,
or whether burning of the even more abundant element O is also required.
Figure~\ref{fig_sum_cn_m5}
shows the sum of the C and N abundance as a function of the C
abundance of the sample of M15 subgiants.  The solid dot shows the
predicted location assuming the initial C and N  abundances
(C$_0$, N$_0$) are the Solar values reduced by the
metallicity of M15 ([Fe/H] = $-2.2$ dex).
Thus this is the initial location for no burning and for a Solar C/N
ratio.  If the present stars incorporated material in which just C was
burned into N, then the locus of the observed
points representing the M15 sample of low luminosity stars should consist of a
single horizontal line, with the initial point, the
presence of no CN-cycle exposed material, at the right end of the line
(the maximum C abundance) and the left end of the line corresponding
to a substantial fraction of the star's mass
(i.e. the atmosphere plus surface convection zone) 
including C-poor, N-rich material.
Furthermore, if the initial C/N ratio of the cluster is not
Solar, then the locus should still be a horizontal line, but located
at a different vertical height in this figure.

The maximum possible N enhancement for a cluster SGB star with these
assumptions occurs if the star formed entirely from material
in which
all C has been converted into N.
For initial values (C$_0$, N$_0$) (not expressed as logarithms),
this maximum N enhancement would be (C$_0$ + N$_0$)/N$_0$.
If the initial value was the Solar ratio, C$_0$/N$_0 \sim3.2$,
the resulting maximum N enhancement is a factor of $\sim$4.2,
while for an unrealistic initial C$_0$/N$_0$ of 10, the maximum N enhancement
is a factor of 11.

Now we examine the behavior of the C and N abundances among
the M15 subgiant sample as inferred from our observations.
It is clear that the assumption that the only thing happening is
inclusion of material in which C was burned into N must be incorrect.
The sum of C+N seems to systematically increase 
by a factor of $\sim$5 between the
most C rich star and most C deficient star.
The discussion of the errors, both internal and systematic,
in \S\ref{section_abund} suggests maximum systematic errors
of $-0.2$ dex for log(C/H) and +0.2 for log(N/H).   This is completely
insufficient to explain such a large trend as errors.

Thus the sum of C+N was
{\it{not}} constant as C was burned into N, wherever that might
have occurred.  Furthermore
the observed range in N abundances is very large.  The most 
obvious way to
reproduce this is to include O burning as well as C burning.
If we adopt Solar ratios as our initial values, then a substantial
amount of O burning is required.

Figure~\ref{fig_sum_cn_m5} suggests that the
initial ratio of C/N is close to Solar.  Adopting the Solar value
as the initial C/N ratio, 
we calculate the minimum amount of O which
must be burned at the base of the AGB envelopes to reproduce the 
locus observed
in the figure (under the arguable assumption of the most extreme of
our stars having formed largely from such material - this will, however,
provide us with at least an estimate of the minimum burning required).
We need to produce a N enhancement of at least a factor of 10.
The Solar ratio is C/N/O = 3.2/1/7.6, so
if all the C and 50\% of the O were converted, we have an enhancement
of N of a factor of 8 available to the present stars.
Oxygen is typically found to be overabundant with respect to Fe in
old metal-poor systems (see Mel\'endez, Barbuy \& Spite 2001,
Gratton \etal\ 2001,
Ram\'{\i}rez \& Cohen 2002, and references therein); we assume
[O/Fe] $\sim +0.3$ dex, a typical value.
Then the initial C/N/O ratios will be 3.2/1/15.2.  Note that the 
same amount
of O has to be burned to produce the observed
distribution of C and N abundances,
but in this case it is a considerably
smaller fraction of the initial O.

\section{Comparison With C and N Studies in Other Globular Clusters
\label{section_othergc}}

We have now analyzed four galactic globular clusters covering
a wide range in metallicity, M71, M5, M13 (see GC--CN), 
and the present study of M15.  In each
case, large samples of stars, all well below the luminosity of the
RGB bump, were used.  In M13 and M71, we had large samples
below the main sequence turn off.  In this section we attempt to assemble,
compare, and integrate the
results of these efforts, adding in relevant other work from the
literature.  The following section will seek to interpret these results.

The nearby globular cluster 47 Tuc has been studied in great detail
by \cite{cannon98} (see this paper for references to many earlier studies),
while \cite{briley04a} extend their results by pushing several magnitudes below
the MSTO of this very nearby cluster.  47 Tuc and the four clusters
we have studied (see GC--CN) are the only globular
clusters for which suitable data exists for the low luminosity
range probed here.

In each of these five globular clusters, there are large differences 
in star-to-star
C and N abundances among the low luminosity stars.  
Fig.~\ref{fig_allgc_c} and Fig.~\ref{fig_allgc_n} show histograms for
the samples of stars in each of these five globular clusters of the
derived C and N abundances.  From these we
estimate
the range of variation among these stars for both C and N for each cluster,
ignoring a few obvious outliers in some cases.  
The field star
C and N values for unmixed stars (low luminosity metal-poor field giants,
presumably these are unmixed stars)
from \cite{gratton00} ([C/Fe] $\sim 0.0$, [N/Fe] $\sim -0.1$ dex) 
roughly coincide
with the maximum C/Fe ratio and with the minimum N/Fe ratio.

Our most important new result derives from Fig.~\ref{fig_allgc_c}
and Fig.~\ref{fig_allgc_n}.
This figure clearly shows that the range of the
spread in both C and N is about the same when expressed as
[C/Fe] and [N/Fe] values in each of the five clusters.

An anti-correlation between C and N has been found in each of these
clusters.  This anti-correlation is most easily seen in the metal rich
clusters as there the observational errors are a smaller fraction of the signal.
However, it is seen even in M15.  This anti-correlation takes a particular
form, illustrated in Fig.~\ref{fig_sum_cn_m5}, where 
the [C/Fe] -- [N/Fe] relationship we earlier demonstrated to prevail in M5
(the dashed curve) is superposed on the results presented here for M15.
The length of
the curve covers the full range of our M5 stellar data.  The
agreement of the mean relations we have determined in M5 and in M15, 
both their form and their extent, is very good.

In the metal-rich GCs M71 and 47 Tuc, the CN and the CH indices appears bimodal, with
a preferred high and low value, each varying with luminosity, 
but few stars occupying the middle ground.  However, the more metal poor 
GCs, M5, M13 and M15 show
no sign of bimodality for either the CH or the CN (or NH) line strengths.
While an upper limit to band strengths in the metal-rich clusters might arise
as the bands saturate, the general appearance of the distribution of 
line indices (see, for example, Fig.~4 of Cannon \etal\ 1998) does
not support this as an important mechanism here.  The bimodality more likely
reflect the underlying abundance distributions of C and of N.

By adding in samples of much more luminous stars on the upper giant branch
with C and N abundances from the literature, we have been able to
show that for M13 and for M15, the most luminous stars have a mean C/Fe
ratio about a factor of 3 to 5 lower than than those on the lower giant
branch.  
A similar behavior is present in the large sample of
luminous stars in M92 studied by \cite{carbon82}, where C and N
abundances were assigned to 43 giants with $M_V < +2$,  by \cite{langer86},
and in the more
recent work on C along the lower RGB of M92 by \cite{bellman01}.  
The latter suggest that 
C depletion begins at $M_V = 0.5$ to 1.0 mag, while \cite{zoccali99}
find the RGB bump to be at $M_V \sim0.0$ mag in M15 based on HST CMDs.
Previous studies of the C and N abundances of stars in M5,
summarized
in \cite{cohen02}, combined with our low luminosity
large sample in this globular cluster, suggest that any  difference in the
mean [C/H] between the tip and the base of the RGB in M5 must be less
than 0.3 dex; further observations to refine this are underway.
For M71, we can compare our main sequence C abundances
with those for the 75 RGB giants found by \cite{briley01m71} from DDO photometry.
There is no evidence for any additional C depletion near the RGB tip in M71,
but the uncertainties are large.
The situation in 47 Tuc is comical.  This globular cluster is so nearby,
hence its stars are relatively bright,
that large surveys of the luminous giants
(see, e.g. Norris, Freeman \& Da Costa 1984) were carried out before the
development of molecular band synthesis techniques, and hence they
compared CH and CN molecular band indices as a function of
luminosity, but did not derive C and N abundances.
We have been unable to put together a sample from the literature
of stars in 47 Tuc with C abundances 
that encompasses the
necessary luminosity range in spite of the multitude of published
analyses.  

This drop in C/Fe near the RGB tip is in agreement with the behavior of field stars
\citep{gratton00}.  Thus, as was suggested in \cite{briley02},
there appear to be two distinct abundance
altering mechanisms that affect globular cluster stars.  One
produces strong star-to-star scatter at all luminosities, and another,
most effective in the metal-poor GCs,
produces the drop in C abundance which 
starts at ${\sim}L$(RGB bump).  Only the latter is seen among
field stars, and only the latter is at present understood, by
enhanced (``deep'') mixing at the first dredge up
once stars evolve to luminosities exceeding  $L$(RGB bump).

\section{Implications for Stellar Evolution \label{section_mix} }

In the previous section, we reviewed the observational results accumulated thus
far regarding the C and N abundances of stars in galactic globular clusters.
We note that the sample of well studied GCs covers a wide range in 
metallicity, and, especially at the metal rich end, a wide range of
present cluster mass and central stellar density. In this section
we discuss the implications of the collected observational results
for stellar evolution of metal-poor globular cluster stars. 
We found that there
appear to be two distinct mixing mechanisms.  First we address the
extra C-depletion found on the upper RGB beginning approximately at $L$(RGB bump).
This mechanism is, to first order at least, understood.

A classical review of post-main sequence stellar evolution can be
found in \cite{iben83}.  Their description of the consequences
of the first dredge up phase, the only dredge up phase to
occur prior to the He flash,
indicates that a doubling of the surface $^{14}$N and a 30\% reduction
in the surface $^{12}$C can be expected, together with a
drop in the ratio of
$^{12}$C/$^{13}$C from the solar value of 89 to $\sim$20,
as well as a drop in surface Li and B by several orders of magnitude.
Observations of metal-poor field stars over a wide range of luminosities
conform fairly well to this picture, see e.g. Shetrone \etal\ (1993),
\cite{gratton00}, although additional mixing of Li and
lower than predicted ratios of $^{12}$C/$^{13}$C seem to 
occur even among field stars \citep{nascimento00}.

Additional physics was introduced into
calculations of dredge up in old metal poor stars to better reproduce
the observations via ``deep mixing''.  Specific improvements include
meridional mixing as described by Sweigart \& Mengel (1979)
as well as turbulent diffusion (see Charbonnel 1994, 1995)
and the insights of Denissenkov \& Denissenkova (1990)
concerning the importance
of the $^{22}$Ne($p,\gamma)^{23}$Na reaction as a way to produce
p-burning nuclei.  
The clear prediction of the most recent calculations of this type
by Denissenkov \& Weiss (1996), Cavallo, Sweigart \&  Bell (1998)
and Weiss, Denissenkov \& Charbonnel (2000) is that
the earliest that deep mixing can
begin is at the location of the bump in the luminosity function
of the RGB which occurs when the H-burning shell crosses a sharp 
molecular weight discontinuity. 

The observations of C and N abundances in globular cluster stars 
(and in field stars, see, e.g. Gratton \etal\ 2000) are
in reasonable agreement with the predictions of the latest
such models with regard to the key points: at what luminosity the
first dredge up begins, the amplitude of the decline in C abundance, and
the general shape of the C depletion as a function of luminosity.
The observational situation is not yet adequate to verify the predicted
increase of the N abundance in GC RGB stars above the bump luminosity.
The models also predict, in agreement with the observations, 
that this phenomenon is more efficient
at low metallicities, as the thickness of the H-burning shell decreases
rapidly  and the shell burning
timescale also decreases as the metallicity rises 
\citep*[see Fig.~7 and 6b of][]{cavallo98}. The models of \cite{cavallo98} and others
can predict the  strong O-Na correlation seen among giants close
to the RGB tip in some globular clusters, particularly the metal poor ones,
as another consequence  of p-burning in the H-burning shell. 

We therefore regard this aspect of the behavior of the 
C and N abundances in globular
cluster stars as having a reasonable explanation.  It is the
very large star-to-star differences in C and N abundances
found at all luminosities in all globular clusters studied 
with sufficient data to date which are not easily explained.
The range of variation of [C/Fe] and of [N/Fe] is to first order
constant, irrespective of the metallicity of the GC and independent
of stellar luminosity, with C and N
anti-correlated, and with maximum C depletions of a factor of
3 to 5 accompanying maximum N enhancement of more than a factor of 10.
The very high N enhancements require
not only CN burning but ON burning as well.
The range of luminosity over which these C and N
variations (and also O and Na variations as well, see, e.g. Cohen \& Melendez 2005
or \cite{gratton01}
and references therein)
occur in globular clusters has by now ruled out
any scenario which invokes dredge up and mixing intrinsic to the star itself.
We must now regard the fundamental origin of the star-to-star
variations in C and N abundance we see in GCs as arising outside the stars whose
spectra we have studied here.

The strong anti-correlation between C and N, however, does
suggest that  CN-cycle material must be involved, and that
this material has somehow reached the surface
of these low luminosity GC stars.   Since we know it cannot come
from inside these stars, it must come from some external source.
As reviewed by Lattanzio, Charbonnel \& Forestini (1999),
CN and ON cycling is known
to occur in intermediate mass AGB stars, and such stars are also known to have sufficient
dredge up to bring such material to their surfaces.   Recent detailed
computations, including both nucleosynthesis with a large set of
isotopes and nuclear reaction pathways, have been carried out by \cite{karakas03}
for very metal poor intermediate mass stars.
\cite{herwig04} has also added in detailed mixing to the stellar surface. 
These calculations can qualitatively
reproduce essentially all of the observational data.

We thus might speculate 
that the site of the proton exposure could be more rapidly evolving
higher mass AGB stars, which then suffered extensive mass loss (either
in or outside of binary systems) and
polluted  the generation of lower-mass stars we currently observe,
while the higher mass stars are now defunct.  Considerable effort
to develop this scenario of AGB pollution of the lower mass stars
we observe today has been made by \cite{ventura01}, \cite{dantona02},
and most recently (and most completely) by \cite{fenner04}.

However, the recent observational facts summarized above have in our view 
rendered this scenario not viable either.  One problem with any ``pollution''
scenario is that these
abundance inhomogeneities cannot simply be surface contaminations
as they would be diluted by the increasing depth of the convective 
envelope during RGB ascent.  Thus  the amount
of accreted mass required to explain the observed C and N variations
becomes very large.  It must be significant fraction of the total stellar mass,
given that a fraction of the star much larger than just the surface
convection zone of a luminous RGB star must be 
contaminated to maintain a constant range of C and of N
at all luminosities.  This seems to us
unrealistic and contrived, but
we note the recent calculations of
\cite{thoul02} which demonstrate that large accumulations
may be possible with some assumptions about stellar orbits,
particularly in clusters with small core radii.  They estimate
that for stars in the core of 47 Tuc as much as 80\% of the 
mass of a 1 M\subsun ~ star in that cluster could be accreted material.
Unfortunately, the core radii of some of the GCs we have studied
are considerably larger.  Even with generous assumptions regarding 
orbital anisotropy, similar
calculations for them yield a much smaller expected accretion ($\sim$10\%), 
which is not sufficient for present purposes. 

Another problem arises because
of the tight anti-correlation between C and N (see Fig.~\ref{fig_sum_cn_m5}).
Any external mechanism for producing these variations 
will involve an efficiency
factor for the incorporation of material.  This might be a cross section
if accretion from the cluster gas is involved, or some property of
the accretion disk if binaries are involved.  We expect this factor
to depend on the mass of the star itself,
how much additional mass is incorporated (${\Delta}M$),  
and the initial C and N abundances in the star itself and within ${\Delta}M$.
Since these properties of
${\Delta}M$ might be expected to fluctuate wildly depending on the
mass of the evolved star producing the N rich ejecta, this 
process therefore should show
a lot of stochastic random variability. ``Pollution'' of a low mass
star by ejecta from
intermediate mass AGB stars is just too
chaotic and unpredictable to be able to reproduce such
well behaved trends.

We now turn to the implications of Fig.~\ref{fig_allgc_c}
and \ref{fig_allgc_n}, which display
the ranges of [C/Fe] and [N/Fe] found among low luminosity
stars (stars at the base of the RGB, subgiants, and/or on the main sequence)
in five GCs spanning a range in metallicity of a factor of 40.  This
figure illustrates our most important new result.  We see
that the maximum and minimum for each of these is approximately the same
(to within a factor of 3)
for each of the clusters.  Note that the maxima and minima
expressed as 
log[$\epsilon$(C)] and log[$\epsilon$(N)] are {\it{not}} constant. 
Furthermore the maximum in [C/Fe] 
corresponds reasonably well to that of the field stars, while the minimum
in [N/Fe] corresponds to that of the field stars.

Thus it seems reasonable that, as is commonly assumed, 
the high C, low N stars represent the nominal
chemical inventory, while the abnormal ones are those with low C and high N.
$^{12}$C is produced by the triple-$\alpha$ process and destroyed by CN
burning, while $^{14}$N is produced via  CN  and ``hot bottom'' burning.
We therefore expect N to behave as a primary element, while the behavior of C may
be more complex.
The observations, however, demand that
the additional material dumped onto the low C/high N stars
is not from some primary process in which
a fixed amount of N per gm is produced, dispersed into
the GC, and mixed into the GC gas.  Instead 
the chemical inventory of C and N
behaves like a secondary process, increasing as [Fe/H] increases.

The modeling of the production and dredge up of elements such as C and N in AGB stars,
while still very uncertain, is rapidly advancing.  Detailed models,
such as those of \cite{karakas03} of yields and of abundances
for various species at the surfaces of such stars after dredge up are now available.
So we must ask what are the surface ratios of the species of interest in 
intermediate mass AGB stars after 
dredge up and what are the relevant mass loss
rates, which will drive the processed material into the cluster gas.
Mass loss in AGB stars is primarily driven
by radiation pressure on dust grains \citep{vasil}.  While there is still some
uncertainty, recent comparisons of heavily obscured AGB stars in the
SMC, the LMC and the Milky Way by \cite{vanloon} suggest that the total mass loss rate
for a star of fixed luminosity is only weakly dependent on metallicity.

Chemical yields for intermediate mass stars have been given by
\cite{marigo}, \cite{ventura02} and \cite{karakas03}.
We examine the ratio of initial to final C and N in models of different initial
metallicity to determine how closely the behavior of
$^{12}$C and of $^{14}$N in the ejecta of such stars matches the
extreme range of the observations in globular clusters over
a wide range in metallicity. 
Table~\ref{table_modelcomp} presents this comparison in detail for the 
models of \cite{ventura02}; those of \cite{karakas03}
and of \cite{gavilan04} do not cover a sufficient
range in metallicity to be useful for our purpose, while the results of 
\cite{marigo} were not given in tabular form.

We find that, while correct in sign, the models of \cite{ventura02}
fail to reproduce the observations by factors of up $\sim$10, depending
on which of the three ratios presented in the table is examined.
The worst discrepancy is in the final ratio C/N after dredge up in the most metal
poor GCs.  There the extent of both the depletion of C 
and the enhancement of N are badly underestimated
by the models.  \cite{fenner04} also finds
that ejecta from intermediate mass AGB stars cannot reproduce details
of the abundance distributions of 
the Mg isotope ratios in NGC~6752 \citep*[see also][]{denissenkov03}.  
AGB ejecta can be observed directly by studying them in situ, i.e. in 
planetary nebulae.
An independent verification 
comparison for the predicted yields
can be attempted via abundance analyses of the planetary nebulae 
in the SMC. \cite{lmc_pn} discuss 
the  planetary nebulae in the LMC, whose metallicity is higher
than that of any GC considered here.
The SMC PN are not as well studied, and
with the demise of STIS/HST, future progress in this area will be at best
very slow.

Although the predictions of \cite{ventura02} and of \cite{karakas03} for
the behavior of C and N after dredge up of intermediate mass AGB stars
are reasonably consistent with each other, the uncertainty in these calculations must
be large.   Whether it is large enough to accommodate discrepancies of
a factor of $\sim$10 in C/N ratios is not clear.
An optimist would say that this level of (dis)-agreement 
is satisfactory, given the difficulties
and complexity of the modeling effort required, while a pessimist would
say that these discrepancies are larger than can be reasonably
expected from the models and from the data. A very recent paper, \cite{ventura05},
discusses the modeling uncertainties arising from just one issue, the
description of convection adopted.  They find changes of a factor of
two in predicted C and N surface abundances in intermediate mass metal poor AGB stars
are easily achieved by this means.  We choose to be optimists.
More such modeling efforts, even though they
require many assumptions,  will be very
valuable.


\section{Implications for Globular Cluster Formation and Chemical Evolution 
\label{section_chem_evol} }

In addition to the accumulated evidence regarding C and N abundances presented
above, there is one other key fact that must figure in any model of the
chemical evolution of globular clusters.  This is that the abundances
of the heavy elements, particularly those between Ca and the Fe-peak, are
constant for all stars in a globular cluster.  Extensive efforts
(see, for example, Cohen \& Melendez 2005) have failed to detect any dispersion
larger than the observational errors.  The abundance spreads are confined
to the light elements\footnote{There is increasing evidence there may be
some variation of the heavy and very rare $r$ and $s$-process elements in
globular clusters.  We ignore this here.}.

We suggest that a viable scenario for the chemical evolution of GCs
can only be constructed if
globular cluster stars are not
all coeval, and more than one epoch of
star formation in GCs must have occurred, albeit all within a relatively
short timescale.  During the early stages,
a variation in C and N abundances satisfying the above observational
data was imprinted on the proto-cluster
gas
{\it{before}} the present generation of stars we now observe were formed.
The low mass stars we currently observe formed from the ``polluted'' gas
some time later during the extended period of star formation in GCs.  Furthermore,
if one is a pessimist, one must rule out 
some previous generation intermediate mass AGB stars as the
the source of this ``pollution''  because of problems in the predicted C/N
ratios.  If one is an optimist, then one ascribes these problems to modeling
difficulties, mass loss 
rates increase dramatically with increasing metallicity, which is not
supported by observations \citep{vanloon}, or some other such factor,
and assumes that these AGB stars did generate the C and N variations seen
today in low luminosity GC stars.

A tentative scenario which fits most of the facts might be that the
first stars to form in the proto-cluster gas were very massive.  Since
this gas might have had very low metallicity, theoretical support
for an IMF heavily biased  towards high mass stars under these conditions
can be found in the review of \cite{bromm04}.
The SNII from these
stars produced the heavy elements through the Fe-peak seen in globular
cluster stars.  The violent explosions ejected energy into the cluster
gas which kept it well mixed.  This is crucial to maintaining constant
abundances of the heavy elements in the stars within a particular GC.  
SNII explosions may also have acted to
disrupt the lowest mass proto-clusters, which became halo field stars.

The lifetimes of high mass stars are very short, and so would be
the duration of this phase of evolution of the GC.
After some (short) time, no more massive stars were formed.
Intermediate mass stars  began to form with metallicity that of the
GC as seen today.  Such stars have typical lifetimes of $\sim$2 Gyr.  
During the course of their evolution, in their
interiors they produced
material that went through the CN process (and the ON process to some extent).
This material was subsequently ejected, but the gas was no longer
mixed globally over the cluster volume, and local pockets of 
substantial or negligible enrichment of the 
light elements developed. Since GC CMD diagrams do not permit
an age range of 2 Gyr among the low mass GC stars, no low mass
stars could have formed until the near the end of this second phase. 
At this point, the low
mass stars that we see today formed, with variable light element ratios,
but fixed heavy element abundances.  

It is now possible to include the formation of globular clusters
in cosmological simulation \citep{kravtsov04}.  However, the level of
detail needed here to follow their chemical evolution with
regard to the light elements is still beyond our capabilities. 
Although the overall picture sketched above seems reasonable, current 
models for nucleosynthesis and dredge up
for intermediate mass AGB stars fail to reproduce in detail
the observed C and N variations in GC stars, in the sense that the 
C depletions and N enhancements observed in low metallicity globular clusters 
are considerably
larger than theory predicts.  Unless those models are flawed,
the relatively short lived stellar source for the second phase of this
scenario, when the cluster gas is no longer well mixed throughout its volume,
is unknown.
  
In evaluating such a scenario, it is important to remember that the
present mass of a GC may be much lower than its initial mass as a proto-cluster;
stars are lost from the cluster through many processes \citep*[see, e.g.][]{mash}.  
Thus the absence
of a relation between the present mass (or central density) of a GC and its
[Fe/H] should not be surprising.
It would be of interest to test to even greater accuracy the
constancy of the Fe-peak elements within a particular GC.

\section{Summary}

We present moderate resolution spectroscopy and photometry for a large sample of 
subgiants and stars 
at the base of the RGB in the extremely metal poor Galactic globular cluster 
M15 (NGC~7078), with the
goal of deriving C abundances (from the G band of CH) and
N abundances (from the NH band at 3360\,\AA). Star-to-star
stochastic variations with significant range
in both [C/Fe] and especially [N/Fe] are found at all luminosities
extending to the subgiants at $M_V {\sim}+3$.

An analysis of these LRIS/Keck spectra with theoretical synthetic
spectra reveals
that these star-to-star variations between C and N 
abundances are anti-correlated, as
would be expected from the presence of proton-capture exposed 
material in
our sample stars. 
The evolutionary states of these stars are such that the 
currently proposed
mechanisms for {\it in situ} modifications of C, N, O, etc. have 
yet to take place.  On this basis,
we infer that the source of proton exposure lies not within the 
present
stars, but more likely in a population of more 
massive  stars which have ``polluted'' our sample. 

The range of variation
of the N abundances is very large and the sum of C+N increases
as C decreases.  To reproduce this requires  
the incorporation not only of CN but also of ON-processed material,
as we also found earlier for  M5 (see GC--CN).

We combine our work with that of \cite{trefzger83} for the brighter
giants in M15 to extend coverage to a larger luminosity
range reaching from the RGB tip to the main sequence turnoff.
We then find strong evidence for additional
depletion of C among the most luminous giants.   This presumably
represents the first dredge up (with enhanced deep mixing) expected for 
such luminous RGB stars in the course of normal
stellar evolution  as they cross the RGB bump.  

Our work now covers four GCs (M15, M13, M5 and M71, see GC--CN)  spanning a metallicity
range of a factor of 40.  We look at the trends of C and N abundances 
common to all the GCs
studied to date, including (from the literature) 47 Tuc.  While all clusters
studied show strong anti-correlated variations of C and N at all
luminosities probed,  the metal rich
clusters (M71, 47 Tuc and M5)
do not show evidence for the first dredge up among their most
luminous giants, while the metal poor ones (M5, M13, M92 and M15)
do.  This is predicted by the models of the first dredge up on the RGB,
which can reproduce essentially all the key features of the associated
changes in C abundance, including the luminosity at which it begins and the amplitude
of the decline in [C/Fe] as a function of metallicity.
The metal poor clusters do not show evidence for the bimodality
in CH and CN line strengths seen in the metal rich clusters.  The
origin of the bimodality is unclear.

It is the star-to-star variations in C and in N seen
at low luminosity in all these GCs that is more difficult to explain.
Having eliminated {\it{in situ}} CN processing, ``pollution'' by material
from intermediate mass AGB stars is the most popular current scenario to
produce this.
However, we rule out this
suggestion, at least as far as accretion onto existing stars is concerned.

\begin{figure}
\epsscale{1.0}
\plotone{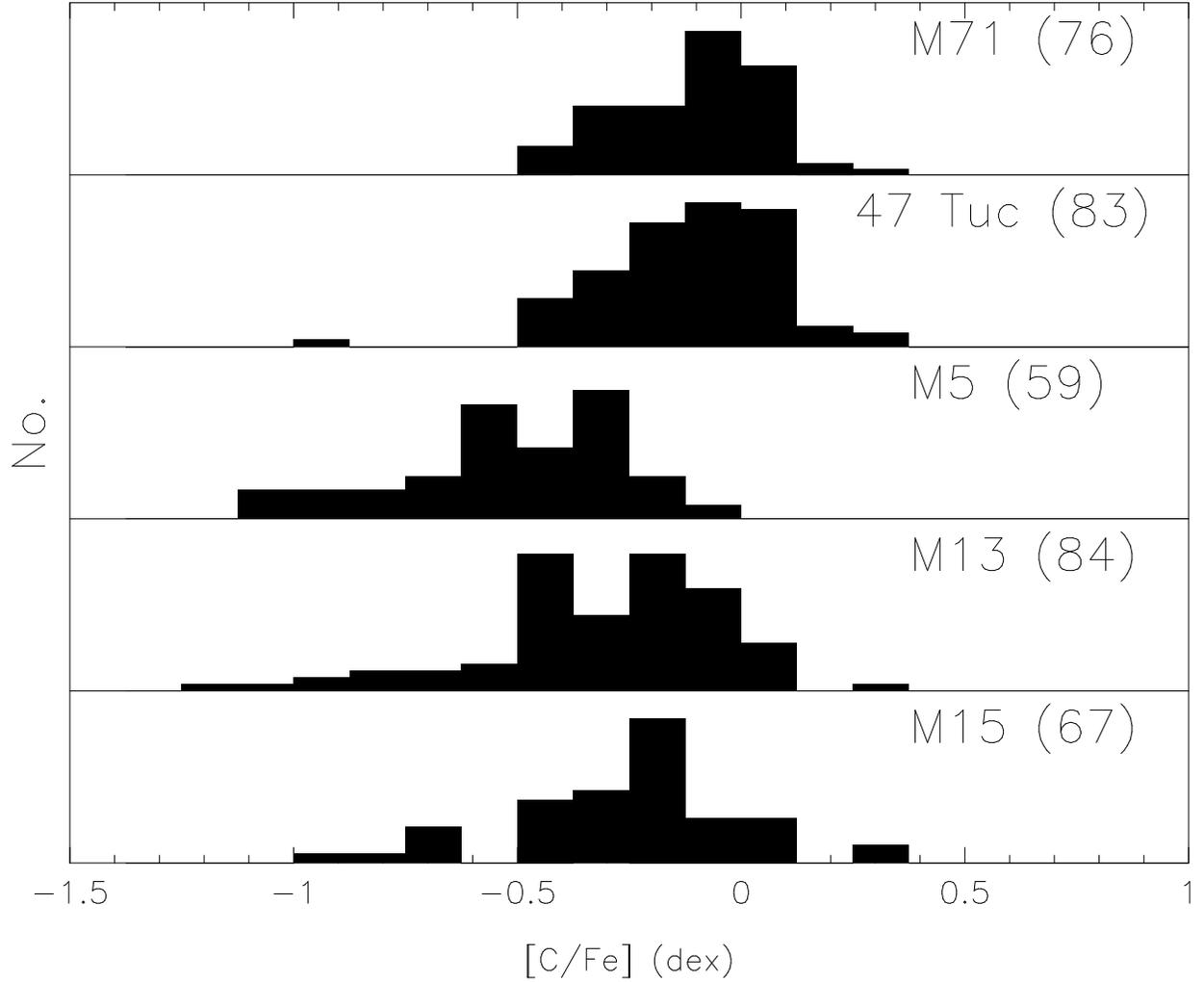}
\caption[]{A histogram of [C/Fe] is shown for  
for the low luminosity stars with suitable
data in each of the globular clusters M15, M13, M5 and M71
from our work (see GC--CN),  as well as 
for 47 Tuc (from Briley et al 2004a).
The region of the main sequence turnoff is excluded as the stars there
have higher \teff\ with very weak molecular bands.
 The total number
of stars included for each cluster is given in parentheses following the
cluster name. 
\label{fig_allgc_c}}
\end{figure}

\begin{figure}
\epsscale{1.0}
\plotone{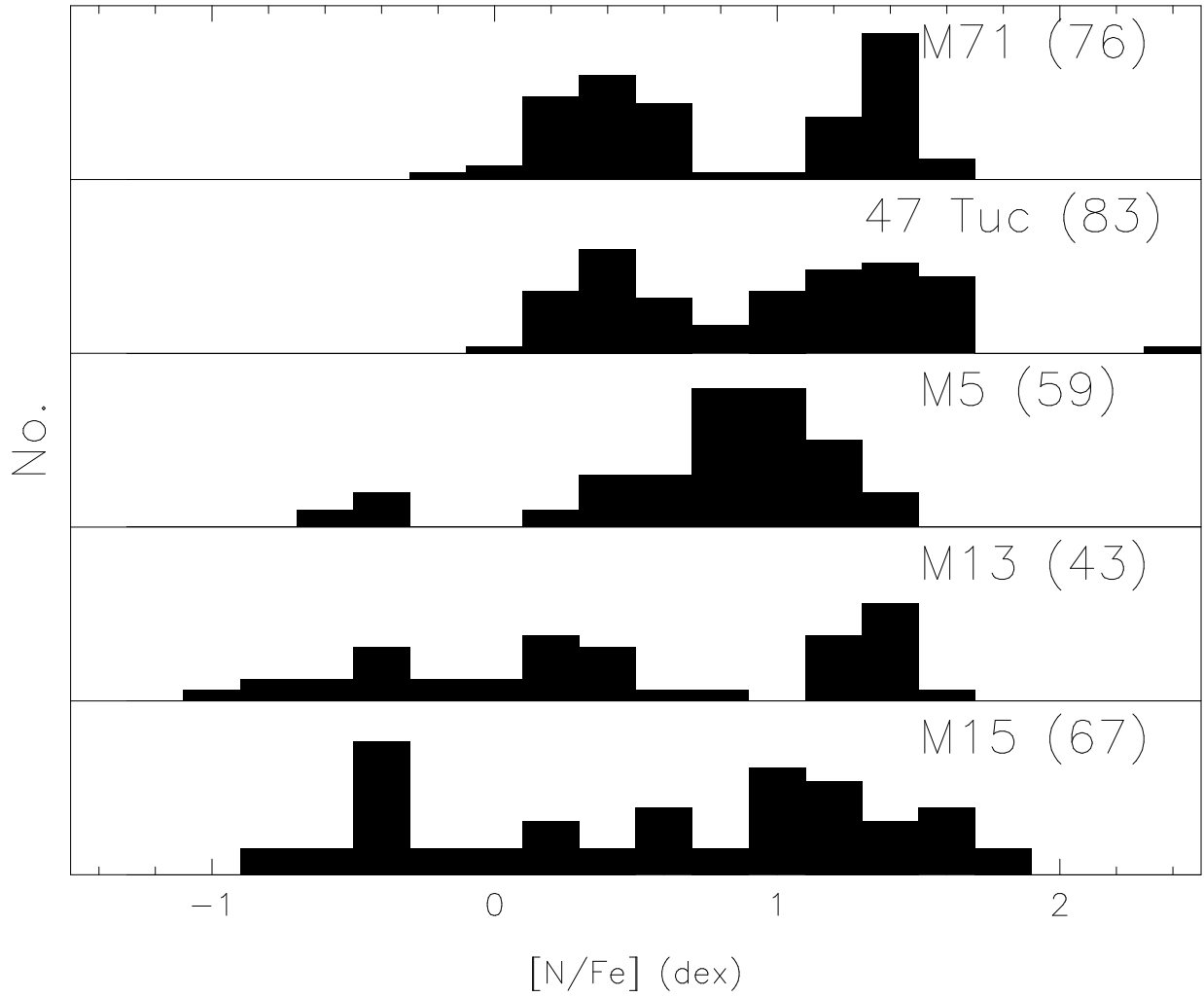}
\caption[]{The same as Fig.~\ref{fig_allgc_c} for [N/Fe].
\label{fig_allgc_n}}
\end{figure}

Our most important new result is that the range of [C/Fe] and of [N/Fe]
seen in these five GCs is approximately constant 
(see Fig.~\ref{fig_allgc_c} and \ref{fig_allgc_n}), 
i.e. C and N are
behaving as though they were produced via a secondary, not a primary, 
nucleosynthesis process.  A detailed comparison of our results with
the models for nucleosynthesis and dredge up of low metallicity intermediate
mass AGB stars by \cite{ventura02} fails to explain the details of the
C and N abundances, predicting key ratios incorrectly by a factor of up to 
$\sim$10.  Thus pollution of
cluster gas by such stars can also be ruled out unless current models
of surface N abundances after dredge up are flawed, which seems possible
given the complexity of the modeling and the many assumptions required.

The behavior of the C and N abundances among low luminosity stars in GCs,
while [Fe/H] is constant to high precision within each GC,
force us to assume that there was an extended period of star 
formation in GCs.  The first stars were exclusively of high mass, and their
SNII ejecta produced the heavy metals seen in the GC. A
second  generation of short-lived stars
of an unknown type (not intermediate mass AGB stars, unless current models
are flawed)
evolved, ejected mass, and ``polluted'' with light elements the GC gas;
the low mass stars we see today formed afterwards.

\acknowledgements

The entire Keck/HIRES and LRIS user communities owes a huge debt to
Jerry Nelson, Gerry Smith, Steve Vogt, Bev Oke, and many other
people who have worked to make the
Keck Telescope and HIRES and LRIS a reality and to operate and
maintain the Keck Observatory. We are grateful to the
W. M.  Keck Foundation for the vision to fund
the construction of the W. M. Keck Observatory.  The authors wish 
to extend
special thanks to those of Hawaiian ancestry on whose sacred mountain
we are privileged to be guests.  Without their generous hospitality,
none of the observations presented herein would
have been possible.

JGC  acknowledges support from the National Science Foundation 
(under grant AST-025951) and
MMB acknowledges support from the National Science Foundation 
(under grant
AST-0098489) and from the F. John Barlow endowed professorship. We are
also in debt to Roger Bell for the use of the SSG program and the 
Dean of the
UW Oshkosh College of Letters and Sciences for the workstation 
which made the extensive modeling possible, and to Jorge Melendez for
the IR observations.

This work has made use of the USNOFS Image and Catalog Archive 
operated by
the United States Naval Observatory, Flagstaff Station
(http://www.nofs.navy.mil/data/fchpix/).
This publication makes use of data from the Two Micron All-Sky Survey,
which is a joint project of the University of Massachusetts and the 
Infrared Processing and Analysis Center, funded by the 
National Aeronautics and Space Administration and the
National Science Foundation.

\clearpage

\clearpage

\clearpage

\clearpage

\clearpage

%
%
\begin{deluxetable}{lllll}
\tablenum{1}
\tablewidth{0pt}
\tablecaption{Photometry for the M15 Sample\label{table_phot}}
\tablehead{
\colhead{ID\tablenotemark{a}} &  \colhead{$V$} &
\colhead{$I$} & \colhead{$J$} & \colhead{$K$} \\
\colhead{}
& \colhead{(mag)} & \colhead{(mag)} & \colhead{(mag)} & 
\colhead{(mag)\tablenotemark{c}}
}
\startdata
C29363\_0823 & 18.03 & 17.17 & 16.61 & 16.05 \\   
C29387\_0716 & 18.27 & 17.40 & \nodata & \nodata \\   
C29388\_0634 & 16.74 & 15.77 & 15.16 & 14.58 \\ 
C29396\_0843 & 17.51 & 16.61 & 16.01 & 15.49 \\ 
C29401\_0908 & 17.70 & 16.81 & 16.21 & 15.68 \\
C29408\_1003 & 18.00 & 17.15 & 16.52 & 16.07  \\
C29413\_1023 & 17.32 & 16.46 & 15.81 & 15.34 \\ 
C29417\_0953 & 18.32 & 17.50 & 16.08 & 16.41 \\  
C29413\_1021 & 18.92 & 18.33 & 14.10 & 13.55  \\
C29424\_0729\tablenotemark{e} & 17.83 & 17.97 &  \nodata & \nodata \\ 
C29426\_0727 & 16.99 & 16.03 & 15.46 & 14.92 \\ 
C29429\_0801 & 16.86 & 15.94 & 15.37 & 14.84 \\
C29442\_0710 & 17.32 & 16.41 & 15.86 & 15.37 \\    
C29443\_0958 & 20.83 & 20.00 &   \nodata & \nodata \\  
C29444\_0618 & 18.60 & 17.85 & 17.34 & 16.91 \\ 
C29445\_0952\tablenotemark{f} & 16.88 & 15.90 & 15.16 & 14.63 \\
C29448\_0655 & 19.51 & \nodata  &   \nodata & \nodata \\
C29448\_0824 & 16.52 & 15.53 & 14.88 & 14.33 \\
C29455\_0524 & 18.60 & 17.78 & 17.28 & \nodata \\    
C29457\_0518 & 19.27 & 18.68 & \nodata & \nodata \\
C29461\_0905 & 19.13 & 18.60 & \nodata & 18.13 \\
C29464\_0644 & 16.93 & 15.98 & 15.42 & 14.90 \\   
C29465\_0859 & 16.91 & 15.89 & 15.26 & 14.72 \\
C29475\_0512 & 18.52 & 17.71 & 17.21 & 16.83 \\
C29485\_0436 & 18.25 & 17.33 & 16.76 & 16.25 \\        
C29485\_0638\tablenotemark{d}
            & 17.35 & 16.50 & 15.95 & 15.50 \\
C29491\_0504 & 17.51 & 16.62 & 15.97 & 15.46 \\  
C29504\_0630 & 18.41 & 17.67 & 17.21 & 16.72 \\ 
C29523\_0432 & 16.98 & 16.03 & 15.35 & 14.80 \\            
C29533\_0714 & 19.26 & 18.66 & \nodata & \nodata \\
C29534\_0711 & 17.72 & 16.84 & 16.27 & 15.77 \\  
C29549\_0416 & 17.86 & 16.97 & 16.27 & 15.76 \\   
C29560\_0528 & 18.42 & 17.60 & 16.97 & 16.46 \\ 
C29573\_0454 & 17.60 & 16.66 & 16.00 & 15.51 \\  
C29582\_0433 & 16.61 & 15.65 & 14.89 & 14.34  \\
C30029\_1248          &  17.07 &  16.17 &  15.46 &  14.94 \\
C30048\_1355          &  18.17 &  17.24 &  16.63 &  16.17 \\
C30051\_1331          &  18.26 &  17.39 &  16.72 &  16.24 \\
C30054\_1253          &  16.67 &  15.73 &  15.02 &  14.49 \\
C30056\_1227          &  18.40 &  17.59 &  17.06 &  16.63 \\
C30069\_1254          &  17.55 &  16.64 &  15.91 &  15.39 \\
C30076\_1230          &  17.46 &  16.56 &  15.88 &  15.35 \\
C30093\_1130          &  18.09 &  17.25 &  16.58 &  16.15 \\
C30095\_1227          &  18.21 &  17.51 & \nodata &  16.16 \\
C30096\_1213          &  16.92 &  16.00 &  15.26 &  14.70 \\
C30102\_1201          &  17.75 &  16.90 &  16.21 &  15.74 (0.05) \\
C30108\_0858          &  16.92 &  15.97 &  15.39 &  14.81 \\
C30108\_1109          &  18.47 &  17.73 &  17.24 &  16.91 \\
C30114\_1055          &  18.18 &  17.34 &  16.86 &  16.35 \\
C30116\_1015          &  17.53 &  16.63 &  16.05 &  15.51 \\
C30116\_1219          &  16.69 &  15.79 &  15.09 &  14.55 \\
C30123\_1138          &  18.27 &  17.41 &  16.89 &  16.38 (0.06) \\
C30124\_0922          &  18.39 &  17.67 &  17.26 &  16.84 \\
C30128\_0840\tablenotemark{d}  
         &  17.85 & \nodata & 17.63 & 17.65 (0.30) \\
C30129\_0946          &  18.05 &  17.22 &  16.61 &  16.08 \\         
C30130\_0945          &  19.53 &  19.00 &  \nodata & \nodata \\
C30130\_1008          &  18.10 &  17.25 &  16.61 &  16.16 \\
C30131\_0800          &  17.02 &  16.11 &  15.44 &  14.86 \\
C30131\_0829          &  16.70 &  15.79 &  15.10 &  14.50 \\
C30141\_0819          &  18.17 &  17.33 &  16.72 &  16.18 \\
C30146\_0829          &  18.28 &  17.47 &  16.85 &  16.35 \\
C30148\_0825          &  15.37 &  14.37 &  13.62 &  12.97 \\
C30149\_1010          &  18.74 &  \nodata &  17.53 &  16.94 \\
C30159\_1044          &  16.76 &  15.81 &  15.18 &  14.61 \\
C30165\_0939          &  18.47 &  17.67 &  17.09 &  16.56 \\
C30179\_0938          &  18.18 &  17.35 &  16.72 &  16.23 \\
C30199\_0745          &  18.26 &  17.47 &  16.82 &  16.30 \\
C30215\_0807          &  17.25 &  16.39 &  15.68 &  15.09 \\
C30247\_0911          &  16.66 &  \nodata &  15.02 &  14.44 \\
C30251\_0915          &  16.72 &  \nodata &  15.08 &  14.50 \\
Non-members\tablenotemark{b} \\
C29393\_0829 & 18.46 & 17.65 & 17.13 & 16.77  \\
C29419\_0508 & 17.88 & 17.00 & 16.45 & 15.93   \\
C30162\_0738 & 16.64 & 15.87 & \nodata & \nodata   \\
\enddata
\tablenotetext{a}{The star names are derived from their
J2000 coordinates.  Star C12345\_5432 has coordinates
21 12 34.5~~+12 54 32 (J2000).}
\tablenotetext{b}{Spectrum appears too strong lined for M15 membership.}
\tablenotetext{c} {$\sigma(K)$ from multiple measurements 
indicated if it exceeds 0.05 mag.}
\tablenotetext{d} {Crowded.}
\tablenotetext{e} {This star is on the extended BHB.}
\tablenotetext{f} {This object is actually a close pair.}
\end{deluxetable}

\begin{deluxetable}{rrrrrrr rrrrr r}
\tablenum{2a}
\tablecaption{I(CH)(4305~\AA\ Band) From Synthetic Spectra \label{table_modelch} }
\rotate
\tablecolumns{8}
\tablewidth{0pc}
\tablehead{
\colhead{T$_{eff}$} & \colhead{log g} & \colhead{$M_V^0$} & \colhead{[C/Fe]} & \colhead{[C/Fe]} & \colhead{[C/Fe]} & \colhead{[C/Fe]} &
\colhead{[C/Fe]} & \colhead{[C/Fe]} & \colhead{[C/Fe]} & \colhead{[C/Fe]} &
\colhead{[C/Fe]} & \colhead{[C/Fe]} \\
\colhead{} & \colhead{} & \colhead{} & \colhead{$-1.4$} & \colhead{$-1.2$} & \colhead{$-1.0$} & \colhead{$-0.8$} &
\colhead{$-0.6$} & \colhead{$-0.4$} & \colhead{$-0.2$} & \colhead{+0.0} & \colhead{+0.2} &
\colhead{+0.4} \\
\colhead{(K)} & \colhead{(dex)} & \colhead{} & \colhead{(dex)} & \colhead{(dex)} & \colhead{(dex)} & \colhead{(dex)} &
\colhead{(dex)} & \colhead{(dex)} & \colhead{(dex)} & \colhead{(dex)} &
\colhead{(dex)} & \colhead{(dex)} 
}
\startdata
5159 & 2.57 & 1.031 & 0.045 & 0.047 & 0.051 & 0.057 & 0.066 & 0.080 & 0.100 & 0.126 & 0.162 & 0.199 \\
5201 & 2.68 & 1.283 & 0.043 & 0.046 & 0.049 & 0.055 & 0.063 & 0.076 & 0.094 & 0.119 & 0.152 & 0.189 \\
5229 & 2.76 & 1.452 & 0.043 & 0.045 & 0.048 & 0.053 & 0.061 & 0.073 & 0.090 & 0.114 & 0.146 & 0.182 \\
5255 & 2.84 & 1.622 & 0.042 & 0.044 & 0.047 & 0.052 & 0.060 & 0.071 & 0.087 & 0.110 & 0.142 & 0.177 \\
5297 & 2.96 & 1.878 & 0.041 & 0.042 & 0.045 & 0.050 & 0.057 & 0.067 & 0.082 & 0.104 & 0.134 & 0.169 \\
5320 & 3.04 & 2.052 & 0.040 & 0.042 & 0.045 & 0.049 & 0.056 & 0.066 & 0.080 & 0.101 & 0.130 & 0.165 \\
5346 & 3.12 & 2.225 & 0.039 & 0.041 & 0.043 & 0.048 & 0.054 & 0.064 & 0.078 & 0.097 & 0.126 & 0.160 \\
5388 & 3.23 & 2.484 & 0.038 & 0.039 & 0.042 & 0.046 & 0.051 & 0.060 & 0.073 & 0.091 & 0.118 & 0.150 \\
5418 & 3.31 & 2.654 & 0.037 & 0.038 & 0.041 & 0.044 & 0.050 & 0.058 & 0.070 & 0.088 & 0.113 & 0.144 \\
5454 & 3.39 & 2.818 & 0.036 & 0.037 & 0.039 & 0.042 & 0.047 & 0.055 & 0.066 & 0.084 & 0.106 & 0.136 \\
5532 & 3.51 & 3.033 & 0.034 & 0.035 & 0.036 & 0.039 & 0.043 & 0.049 & 0.058 & 0.071 & 0.091 & 0.117 \\
5607 & 3.57 & 3.136 & 0.032 & 0.033 & 0.034 & 0.036 & 0.039 & 0.043 & 0.050 & 0.061 & 0.078 & 0.099 \\
5739 & 3.66 & 3.229 & 0.029 & 0.029 & 0.030 & 0.031 & 0.033 & 0.036 & 0.040 & 0.047 & 0.059 & 0.074 \\
5976 & 3.78 & 3.337 & 0.025 & 0.025 & 0.026 & 0.026 & 0.027 & 0.028 & 0.030 & 0.033 & 0.038 & 0.046 \\
6217 & 3.91 & 3.461 & 0.022 & 0.022 & 0.022 & 0.022 & 0.023 & 0.023 & 0.024 & 0.026 & 0.028 & 0.031 \\
6387 & 4.03 & 3.641 & 0.020 & 0.020 & 0.020 & 0.020 & 0.020 & 0.020 & 0.021 & 0.021 & 0.023 & 0.024 \\
6437 & 4.13 & 3.864 & 0.020 & 0.020 & 0.020 & 0.020 & 0.020 & 0.021 & 0.021 & 0.022 & 0.023 & 0.024 \\
6412 & 4.22 & 4.118 & 0.018 & 0.018 & 0.018 & 0.019 & 0.019 & 0.019 & 0.020 & 0.021 & 0.022 & 0.024 \\
\enddata
\end{deluxetable}

\begin{deluxetable}{rrrrr rrrrr rrrrr}
\tablenum{2b}
\tablecaption{I(NH)(3360~\AA\ Band) From Synthetic Spectra \label{table_modelnh} }
\rotate
\tablecolumns{8}
\tablewidth{0pc}
\tablehead{
\colhead{$M_V^0$} & \colhead{} & \colhead{} & \colhead{} & \colhead{} &
\colhead{} & \multispan{2}{[N/Fe]} \\
\colhead{} & \colhead{$-0.6$} & \colhead{$-0.4$} & \colhead{$-0.2$} & \colhead{+0.0} & 
\colhead{+0.2} & \colhead{+0.4} & \colhead{+0.8} & \colhead{+1.0} & \colhead{+1.2} & 
\colhead{+1.4} & \colhead{+1.6} & \colhead{+1.8} & \colhead{+2.0} \\
\colhead{} & \colhead{} & \colhead{} & \colhead{} & \colhead{} &
\colhead{} & \multispan{2}{(dex)} 
}
\startdata
1.031  &  0.030  &  0.037  &  0.046  &  0.060  &  0.078  &  0.100  &  0.127  &  0.158  &  0.192  &  0.229  &  0.264  &  0.294  &  0.316  &  0.329  \\
1.283  &  0.028  &  0.034  &  0.043  &  0.056  &  0.073  &  0.095  &  0.121  &  0.151  &  0.184  &  0.221  &  0.257  &  0.288  &  0.313  &  0.327  \\
1.452  &  0.027  &  0.033  &  0.042  &  0.054  &  0.070  &  0.091  &  0.117  &  0.147  &  0.180  &  0.216  &  0.252  &  0.285  &  0.310  &  0.326  \\
1.622  &  0.026  &  0.032  &  0.040  &  0.052  &  0.068  &  0.088  &  0.114  &  0.143  &  0.176  &  0.212  &  0.248  &  0.282  &  0.308  &  0.325  \\
1.878  &  0.025  &  0.030  &  0.038  &  0.049  &  0.064  &  0.084  &  0.108  &  0.137  &  0.169  &  0.204  &  0.241  &  0.276  &  0.303  &  0.323  \\
2.052  &  0.024  &  0.029  &  0.037  &  0.047  &  0.062  &  0.082  &  0.106  &  0.134  &  0.167  &  0.201  &  0.238  &  0.273  &  0.301  &  0.322  \\
2.225  &  0.023  &  0.028  &  0.035  &  0.045  &  0.060  &  0.079  &  0.103  &  0.131  &  0.163  &  0.197  &  0.234  &  0.269  &  0.299  &  0.320  \\
2.484  &  0.021  &  0.026  &  0.032  &  0.042  &  0.056  &  0.074  &  0.096  &  0.124  &  0.155  &  0.190  &  0.226  &  0.262  &  0.293  &  0.316  \\
2.654  &  0.020  &  0.024  &  0.030  &  0.040  &  0.053  &  0.070  &  0.092  &  0.119  &  0.150  &  0.184  &  0.220  &  0.257  &  0.289  &  0.313  \\
2.818  &  0.018  &  0.022  &  0.028  &  0.037  &  0.049  &  0.065  &  0.086  &  0.113  &  0.143  &  0.177  &  0.213  &  0.249  &  0.282  &  0.308  \\
3.033  &  0.015  &  0.018  &  0.023  &  0.030  &  0.040  &  0.054  &  0.073  &  0.097  &  0.125  &  0.158  &  0.192  &  0.229  &  0.264  &  0.294  \\
3.136  &  0.012  &  0.015  &  0.018  &  0.024  &  0.032  &  0.044  &  0.060  &  0.081  &  0.106  &  0.137  &  0.170  &  0.206  &  0.242  &  0.275  \\
3.229  &  0.008  &  0.009  &  0.012  &  0.015  &  0.021  &  0.029  &  0.040  &  0.056  &  0.077  &  0.103  &  0.133  &  0.166  &  0.201  &  0.237  \\
3.337  &  0.002  &  0.003  &  0.004  &  0.005  &  0.008  &  0.012  &  0.017  &  0.025  &  0.037  &  0.054  &  0.075  &  0.101  &  0.132  &  0.165  \\
3.461  & -0.002  & -0.002  & -0.002  & -0.001  &  0.000  &  0.002  &  0.004  &  0.008  &  0.014  &  0.023  &  0.035  &  0.052  &  0.074  &  0.101  \\  
3.641  & -0.005  & -0.005  & -0.005  & -0.004  & -0.003  & -0.002  & -0.001  &  0.001  &  0.005  &  0.010  &  0.019  &  0.030  &  0.046  &  0.067  \\
3.864  & -0.006  & -0.006  & -0.005  & -0.005  & -0.004  & -0.003  & -0.002  &  0.000  &  0.003  &  0.008  &  0.016  &  0.027  &  0.042  &  0.062  \\
4.118  & -0.006  & -0.006  & -0.005  & -0.005  & -0.004  & -0.003  & -0.001  &  0.001  &  0.005  &  0.011  &  0.020  &  0.032  &  0.049  &  0.071  \\
\enddata
\end{deluxetable}

\newpage
\begin{deluxetable}{rrrrrrrrrr}
\tablenum{3}
\rotate
\tablecolumns{8}
\tablewidth{0pc}
\tablecaption{Deduced C and N and Abundances for M15 Sample
   \label{table_obs_inds}}
\tablehead{
\colhead{Star} &
\colhead{V} &
\colhead{B$-$V} &
\colhead{M$_V$} &
\colhead{T$_{eff}$} &
\colhead{log g} &
\colhead{CH} &
\colhead{NH} &
\colhead{[C/Fe]} &
\colhead{[N/Fe]} \\
\colhead{} &
\colhead{(mag)} & \colhead{(mga)} & \colhead{(mag)} &
\colhead{(K)} &
\colhead{(dex)} &
\colhead{(\%)} & \colhead{(\%)} &
\colhead{(dex)} & \colhead{(dex)}
}
\startdata
C30148\_0825 & 15.37 & 0.81 & 0.06 & 4939 & 2.05 & 0.123$\pm$0.008 &  0.244$\pm$0.024  & $-0.27^{+0.06}_{-0.06}$ & $1.12^{+0.15}_{-0.14}$ \\
C29448\_0824 & 16.52 & 0.74 & 1.21 & 5143 & 2.60 & 0.106$\pm$0.007 &  0.013$\pm$0.010  & $-0.11^{+0.06}_{-0.06}$ & $\le -0.6$ \\
C29582\_0433 & 16.61 & 0.74 & 1.30 & 5157 & 2.65 & 0.099$\pm$0.008 &  0.187$\pm$0.016  & $-0.15^{+0.06}_{-0.07}$ & $1.02^{+0.09}_{-0.09}$ \\
C30247\_0911 & 16.66 & 0.69 & 1.35 & 5166 & 2.67 & 0.095$\pm$0.006 & $-0.031\pm$0.014  & $-0.17^{+0.06}_{-0.06}$ & $\le -0.6$ \\
C30054\_1253 & 16.67 & 0.67 & 1.36 & 5167 & 2.67 & 0.102$\pm$0.006 &  0.190$\pm$0.018  & $-0.11^{+0.05}_{-0.06}$ & $1.05^{+0.10}_{-0.10}$ \\
C30116\_1219 & 16.69 & 0.74 & 1.38 & 5170 & 2.68 & 0.101$\pm$0.006 &  0.037$\pm$0.017  & $-0.12^{+0.05}_{-0.06}$ & $-0.31^{+0.30}_{-0.66}$ \\
C30131\_0829 & 16.70 & 0.73 & 1.39 & 5172 & 2.68 & 0.103$\pm$0.007 &  0.041$\pm$0.015  & $-0.10^{+0.05}_{-0.06}$ & $-0.23^{+0.25}_{-0.43}$ \\
C29388\_0634 & 16.74 & 0.70 & 1.43 & 5178 & 2.71 & 0.104$\pm$0.004 &  0.104$\pm$0.012  & $-0.08^{+0.03}_{-0.03}$ & $0.50^{+0.09}_{-0.10}$ \\
C30159\_1044 & 16.76 & 0.74 & 1.45 & 5181 & 2.72 & 0.113$\pm$0.008 &  0.115$\pm$0.028  & $-0.01^{+0.05}_{-0.06}$ & $0.59^{+0.19}_{-0.22}$ \\
C29429\_0801 & 16.86 & 0.69 & 1.55 & 5195 & 2.77 & 0.082$\pm$0.005 &  0.028$\pm$0.016  & $-0.27^{+0.06}_{-0.06}$ & $-0.54^{+0.40}_{-0.82}$ \\
C29445\_0952\tablenotemark{b} & 16.88 & 0.75 & 1.57 & 5198 & 2.78 & 0.059$\pm$0.004 &  0.071$\pm$0.024  & \nodata & \nodata \\
C29465\_0859 & 16.91 & 0.68 & 1.60 & 5203 & 2.79 & 0.100$\pm$0.006 &  0.247$\pm$0.015  & $-0.09^{+0.05}_{-0.05}$ & $1.39^{+0.09}_{-0.08}$ \\
C30108\_0858 & 16.92 & 0.71 & 1.61 & 5205 & 2.79 & 0.061$\pm$0.004 &  0.259$\pm$0.017  & $-0.58^{+0.08}_{-0.10}$ & $1.46^{+0.10}_{-0.10}$ \\
C30096\_1213 & 16.92 & 0.63 & 1.61 & 5205 & 2.79 & 0.124$\pm$0.009 &  0.025$\pm$0.019  & $ 0.09^{+0.06}_{-0.06}$ & $\le -0.6$ \\
C29464\_0644 & 16.93 & 0.71 & 1.62 & 5207 & 2.80 & 0.077$\pm$0.006 &  0.213$\pm$0.016  & $-0.32^{+0.07}_{-0.09}$ & $1.21^{+0.09}_{-0.09}$ \\
C29523\_0432 & 16.98 & 0.69 & 1.67 & 5215 & 2.82 & 0.116$\pm$0.009 &  0.066$\pm$0.018  & $ 0.05^{+0.06}_{-0.07}$ & $0.19^{+0.19}_{-0.23}$ \\
C29426\_0727 & 16.99 & 0.69 & 1.68 & 5216 & 2.82 & 0.081$\pm$0.006 &  0.173$\pm$0.014  & $-0.25^{+0.06}_{-0.07}$ & $0.99^{+0.08}_{-0.08}$ \\
C30131\_0800 & 17.02 & 0.72 & 1.71 & 5220 & 2.84 & 0.087$\pm$0.003 & \nodata  & $-0.18^{+0.04}_{-0.05}$ & \nodata \\
C30029\_1248 & 17.07 & 0.68 & 1.76 & 5227 & 2.86 & 0.089$\pm$0.004 &  0.197$\pm$0.021  & $-0.15^{+0.04}_{-0.05}$ & $1.14^{+0.12}_{-0.12}$ \\
C30215\_0807 & 17.25 & 0.65 & 1.94 & 5256 & 2.95 & 0.117$\pm$0.008 &  0.081$\pm$0.017  & $ 0.11^{+0.05}_{-0.06}$ & $0.38^{+0.15}_{-0.17}$ \\
C29442\_0710 & 17.32 & 0.68 & 2.01 & 5266 & 2.97 & 0.056$\pm$0.004 &  0.228$\pm$0.015  & $-0.61^{+0.09}_{-0.10}$ & $1.34^{+0.08}_{-0.08}$ \\
C29413\_1023 & 17.32 & 0.69 & 2.01 & 5266 & 2.97 & 0.154$\pm$0.010 &  0.317$\pm$0.016  & $ 0.33^{+0.06}_{-0.06}$ & $1.94^{+ \cdots}_{-0.15}$ \\
C29485\_0638 & 17.35 & 0.61 & 2.04 & 5270 & 2.99 & 0.065$\pm$0.004 &  0.113$\pm$0.018  & $-0.42^{+0.07}_{-0.08}$ & $0.65^{+0.13}_{-0.14}$ \\
C30076\_1230 & 17.46 & 0.60 & 2.15 & 5288 & 3.04 & 0.083$\pm$0.006 &  0.257$\pm$0.025  & $-0.15^{+0.07}_{-0.08}$ & $1.52^{+0.15}_{-0.14}$ \\
C29491\_0504 & 17.51 & 0.64 & 2.20 & 5295 & 3.06 & 0.098$\pm$0.008 &  0.078$\pm$0.022  & $ 0.00^{+0.07}_{-0.07}$ & $0.39^{+0.19}_{-0.24}$ \\
C29396\_0843 & 17.51 & 0.71 & 2.20 & 5295 & 3.06 & 0.089$\pm$0.002 &  0.008$\pm$0.019  & $-0.080^{+0.02}_{-0.02}$ & $\le -0.60$ \\
C30116\_1015 & 17.53 & 0.71 & 2.22 & 5298 & 3.07 & 0.080$\pm$0.006 &  0.011$\pm$0.027  & $-0.18^{+0.07}_{-0.08}$ & $\le -0.60$ \\
C30069\_1254 & 17.55 & 0.61 & 2.24 & 5301 & 3.08 & 0.067$\pm$0.005 &  0.116$\pm$0.021  & $-0.36^{+0.08}_{-0.08}$ & $0.70^{+0.14}_{-0.16}$ \\
C29573\_0454 & 17.60 & 0.68 & 2.29 & 5310 & 3.10 & 0.115$\pm$0.009 &  0.349$\pm$0.025  & $ 0.14^{+0.06}_{-0.06}$ & $\ge +2.00$ \\
C29401\_0908 & 17.70 & 0.68 & 2.39 & 5327 & 3.15 & 0.062$\pm$0.005 &  0.220$\pm$0.019  & $-0.38^{+0.07}_{-0.07}$ & $1.35^{+0.11}_{-0.11}$ \\
C29534\_0711 & 17.72 & 0.66 & 2.41 & 5330 & 3.16 & 0.070$\pm$0.005 & $-0.046\pm$0.029  & $-0.27^{+0.07}_{-0.08}$ & $\le -0.60$ \\
C30102\_1201 & 17.75 & 0.62 & 2.44 & 5336 & 3.17 & 0.046$\pm$0.004 &  0.243$\pm$0.044  & $-0.81^{+0.19}_{-0.20}$ & $1.49^{+0.26}_{-0.24}$ \\
C29424\_0729\tablenotemark{a} & 17.85 & $-0.08$ & 2.54 & $>10,000$ & \nodata & 0.004$\pm$0.002 &  \nodata  & \nodata & \nodata \\
C29549\_0416 & 17.86 & 0.64 & 2.55 & 5356 & 3.23 & 0.079$\pm$0.006 & $-0.002\pm$0.032  & $-0.12^{+0.07}_{-0.07}$ & $\le -0.6$ \\
C29408\_1003 & 18.00 & 0.67 & 2.69 & 5388 & 3.30 & 0.089$\pm$0.006 &  0.031$\pm$0.018  & $ 0.02^{+0.05}_{-0.06}$ & $-0.17^{+0.32}_{- \cdots}$ \\
C29363\_0823 & 18.03 & 0.67 & 2.72 & 5396 & 3.31 & 0.075$\pm$0.006 &  0.031$\pm$0.024  & $-0.13^{+0.06}_{-0.07}$ & $-0.16^{+0.40}_{- \cdots}$ \\
C30129\_0946 & 18.05 & 0.65 & 2.74 & 5401 & 3.32 & 0.058$\pm$0.005 &  0.297$\pm$0.046  & $-0.36^{+0.09}_{-0.11}$ & $1.88^{+ \cdots}_{-0.30}$ \\
C30093\_1130 & 18.09 & 0.58 & 2.78 & 5413 & 3.34 & 0.069$\pm$0.005 &  0.023$\pm$0.031  & $-0.17^{+0.06}_{-0.08}$ & $-0.38^{+0.64}_{- \cdots}$ \\
C30130\_1008 & 18.10 & 0.68 & 2.79 & 5416 & 3.34 & 0.049$\pm$0.004 &  0.146$\pm$0.035  & $-0.56^{+0.11}_{-0.13}$ & $1.01^{+0.21}_{-0.23}$ \\
C30141\_0819 & 18.17 & 0.65 & 2.86 & 5440 & 3.38 & 0.067$\pm$0.007 &  0.041$\pm$0.049  & $-0.27^{+0.10}_{-0.13}$ & $0.09^{+0.55}_{- \cdots}$ \\
C30048\_1355 & 18.17 & 0.70 & 2.86 & 5440 & 3.38 & 0.061$\pm$0.006 &  0.045$\pm$0.041  & $-0.17^{+0.09}_{-0.11}$ & $0.16^{+0.46}_{- \cdots}$ \\
C30179\_0938 & 18.18 & 0.64 & 2.87 & 5444 & 3.38 & 0.044$\pm$0.004 &  0.173$\pm$0.028  & $-0.69^{+0.13}_{-0.17}$ & $1.19^{+0.16}_{-0.16}$ \\
C30114\_1055 & 18.18 & 0.62 & 2.87 & 5444 & 3.38 & 0.056$\pm$0.005 &  0.151$\pm$0.028  & $-0.36^{+0.09}_{-0.10}$ & $1.07^{+0.16}_{-0.18}$ \\
C30095\_1227 & 18.21 & 0.57 & 2.90 & 5457 & 3.41 & 0.074$\pm$0.006 & $-0.018\pm$0.034  & $-0.06^{+0.06}_{-0.07}$ & $\le -0.6$ \\
C29485\_0436 & 18.25 & 0.65 & 2.94 & 5475 & 3.43 & 0.081$\pm$0.007 &  0.007$\pm$0.032  & $ 0.03^{+0.07}_{-0.07}$ & $\le -0.6$ \\
C30199\_0745 & 18.26 & 0.60 & 2.95 & 5480 & 3.44 & 0.052$\pm$0.004 & $-0.002\pm$0.047  & $-0.41^{+0.09}_{-0.11}$ & $\le -0.6$ \\
C30051\_1331 & 18.26 & 0.70 & 2.95 & 5480 & 3.44 & 0.085$\pm$0.006 &  0.019$\pm$0.035  & $ 0.08^{+0.06}_{-0.06}$ & $-0.43^{+0.77}_{- \cdots}$ \\
C30123\_1138 & 18.27 & 0.57 & 2.96 & 5485 & 3.44 & 0.109$\pm$0.008 &  0.254$\pm$0.034  & $ 0.29^{+0.05}_{-0.06}$ & $1.69^{+0.22}_{-0.19}$ \\
C29387\_0716 & 18.27 & 0.64 & 2.96 & 5485 & 3.44 & 0.043$\pm$0.003 &  0.234$\pm$0.022  & $-0.67^{+0.14}_{-0.14}$ & $1.58^{+0.12}_{-0.12}$ \\
C30146\_0829 & 18.28 & 0.64 & 2.97 & 5491 & 3.45 & 0.058$\pm$0.004 &  0.088$\pm$0.028  & $-0.26^{+0.08}_{-0.08}$ & $0.68^{+0.21}_{-0.26}$ \\
C29417\_0953 & 18.32 & 0.63 & 3.01 & 5515 & 3.47 & 0.061$\pm$0.004 &  0.109$\pm$0.019  & $-0.17^{+0.07}_{-0.08}$ & $0.87^{+0.13}_{-0.14}$ \\
C30124\_0922 & 18.39 & 0.52 & 3.08 & 5574 & 3.52 & 0.024$\pm$0.002 & $-0.010\pm$0.032  & \nodata & \nodata \\
C30056\_1227 & 18.40 & 0.56 & 3.09 & 5585 & 3.53 & 0.024$\pm$0.005 &  0.226$\pm$0.040  & \nodata & $1.64^{+0.23}_{-0.22}$ \\
C29504\_0630 & 18.41 & 0.54 & 3.10 & 5596 & 3.53 & 0.024$\pm$0.003 &  0.099$\pm$0.024  & \nodata & $0.89^{+0.17}_{-0.20}$ \\
C29560\_0528 & 18.42 & 0.62 & 3.11 & 5609 & 3.54 & 0.049$\pm$0.004 &  0.120$\pm$0.042  & $-0.28^{+0.09}_{-0.12}$ & $1.05^{+0.26}_{-0.32}$ \\
C30165\_0939 & 18.47 & 0.58 & 3.16 & 5690 & 3.59 & 0.020$\pm$0.015 &  0.345$\pm$0.059  & \nodata & $\ge +2.0$ \\
C30108\_1109 & 18.47 & 0.54 & 3.16 & 5690 & 3.59 & 0.017$\pm$0.003 &  0.026$\pm$0.031  & \nodata & $0.11^{+0.51}_{- \cdots}$ \\
C29475\_0512 & 18.52 & 0.55 & 3.21 & 5802 & 3.65 & 0.026$\pm$0.003 &  0.131$\pm$0.028  & \nodata & $1.33^{+0.17}_{-0.19}$ \\
C29455\_0524 & 18.60 & 0.56 & 3.29 & 6000 & 3.75 & 0.040$\pm$0.003 &  0.030$\pm$0.030  & $ 0.06^{+0.08}_{-0.12}$ & $0.68^{+0.39}_{- \cdots}$ \\
C29444\_0618 & 18.60 & 0.54 & 3.29 & 6000 & 3.75 & 0.025$\pm$0.003 &  0.081$\pm$0.028  & \nodata & $1.26^{+0.21}_{-0.26}$ \\
C29413\_1021 & 18.92 & 0.45 & 3.61 & 6477 & 4.02 & 0.028$\pm$0.003 &  0.000$\pm$0.024  & \nodata & $0.68^{+0.78}_{- \cdots}$ \\
C29461\_0905 & 19.13 & 0.43 & 3.82 & 6569 & 4.12 & 0.027$\pm$0.002 &  \nodata  & \nodata & \nodata \\
C30149\_1009 & 19.16 & 0.50 & 3.85 & 6575 & 4.14 & 0.028$\pm$0.005 & $-0.010\pm$0.046  & \nodata & \nodata \\
C29533\_0714 & 19.26 & \nodata & 3.95 & 6575 & 4.17 & 0.008$\pm$0.005 &  \nodata  & \nodata & \nodata \\
C29457\_0518 & 19.27 & 0.44 & 3.96 & 6581 & 4.18 & 0.022$\pm$0.003 &  \nodata  & $-0.46^{+0.85}_{- \cdots}$ & \nodata \\
C29448\_0655 & 19.51 & 0.44 & 4.20 & 6559 & 4.27 & 0.010$\pm$0.004 &  \nodata  & \nodata & \nodata \\
C30130\_0945 & 19.53 & 0.45 & 4.22 & 6554 & 4.27 & 0.015$\pm$0.005 & $-0.005\pm$0.049  & \nodata & \nodata \\
C29443\_0958 & 20.83 & 0.60 & 5.52 & 5997 & 4.58 & 0.027$\pm$0.004 & $-0.011\pm$0.038  & \nodata & \nodata
\enddata
\tablenotetext{a}{Hot star on extended HB in M15.}
\tablenotetext{b}{This object is a close pair.}
\end{deluxetable}

\clearpage

%
%
\begin{deluxetable}{ccccccccccccc}\tablecolumns{13}
\tablenum{4}
\tablewidth{0pt}
\rotate
\tablecaption{Changes in Derived C and N Abundances for Different
Model Parameters}
\tablehead{
\colhead{}&\colhead{}&\colhead{}&\multicolumn{2}{c}{$\Delta$(m-
M)$_V$ = --0.10}&\multicolumn{2}{c}{$\Delta$[O/Fe] = +0.20} &
\multicolumn{2}{c}{\ciso = 4\tablenotemark{a}}
   & \multicolumn{2}{c}{[Fe/H] = --2.42\tablenotemark{b}}
   &\multicolumn{2}{c}{Turb = 1.5 km/s} \\
\colhead{Star}&\colhead{[C/Fe]}&\colhead{[N/Fe]}&\colhead{$\Delta$[C/ 
Fe]}&\colhead{
$\Delta$[N/Fe]}&\colhead{$\Delta$[C/Fe]}&\colhead{$\Delta$[N/ 
Fe]}&\colhead{
$\Delta$[C/Fe]}&\colhead{$\Delta$[N/Fe]}
&\colhead{$\Delta$[C/Fe]}&\colhead{$\Delta$[N/Fe]}&\colhead{$\Delta$[C/ 
Fe]
}&\colhead{$\Delta$[N/Fe]}
}
\startdata
C30215\_0807&   0.11 &  0.38 &   0.01 &  0.01 &   0.00 &   0.00 &    
0.00 &   0.00 &   0.14 &   0.16 &   0.02 &   0.00 \\
C29442\_0710& --0.61 &  1.34 &   0.01 &  0.01 & --0.01 &   0.00 &    
0.00 &   0.00 &   0.23 &   0.15 &   0.04 &   0.01 \\
C29387\_0716& --0.67 &  1.58 &   0.12 &  0.07 & --0.01 &   0.00 &  
--0.01 &   0.00 &   0.33 &   0.16 &   0.07 &   0.03 \\
C30146\_0829& --0.26 &  0.68 &   0.11 &  0.09 &   0.00 &   0.00 &    
0.00 &   0.00 &   0.23 &   0.17 &   0.05 &   0.01
\enddata
\tablenotetext{a}{Reduced from the adopted value of \ciso = 10.}
\tablenotetext{b}{Reduced from the adopted value of [Fe/H] = $-2.22$ dex.}
\label{table_changes}
\end{deluxetable}

%
%
\begin{deluxetable}{l ccc | ccc}
\tablenum{5}
\tablewidth{0pt}
\tablecaption{Comparison of C and N Observed Abundances with Models 
of \cite{ventura02} \label{table_modelcomp}}
\tablehead{
\colhead{} & \colhead{} & \colhead{Models} & \colhead{} & 
\colhead{} & \colhead{Obs.} & \colhead{} \\
\colhead{[Fe/H]} &
\colhead{$C_f/C_0$} & \colhead{$N_f/N_0$} & \colhead{[$N_f/C_f$]} &
\colhead{$C_f/C_0$} & \colhead{$N_f/N_0$} & \colhead{[$N_f/C_f$]}  \\
\colhead{} & \colhead{} & \colhead{} & \colhead{(dex)} &
\colhead{} & \colhead{} & \colhead{(dex)} 
}
\startdata
$-2.4$ (M15) &          0.75 & 28 & +1.6   &    0.20 & 100 & +2.7 \\
$-0.8$ (M71, 47 Tuc) &  0.11 & ~7 & +1.8   &    0.25 & 15 & +1.8 \\

\enddata
\end{deluxetable}

\end{document}